\documentclass[12pt,letterpaper]{article}
\usepackage[margin=1in]{geometry}
\usepackage{graphicx} 
\usepackage{amsmath,amssymb,bbm,mathtools,amsthm}
\usepackage{booktabs,diagbox,array,multirow,enumitem}
\usepackage{float}
\usepackage{subcaption}
\usepackage{hyperref}
\usepackage{xcolor}
\usepackage{algorithm}
\usepackage{algpseudocode}
\usepackage[
backend=bibtex,
style=alphabetic,
sorting=ynt,
maxbibnames=99
]{biblatex}
\newtheorem{theorem}{Theorem}[section]

\newtheorem{definition}{Definition}[section]
\newtheorem{lem}[theorem]{Lemma}

\newtheorem{assum}{Assumption}[section]

\newcommand\iidsim{\stackrel{\mathclap{\scriptsize\mbox{iid}}}{\sim}}

\DeclareMathOperator{\Tr}{Tr} 
\DeclareMathOperator{\Id}{Id}

\title{Nonparametric shrinkage estimation of high-dimensional variance component matrices}

\author{%
 Ran Xie\thanks{Department of Statistics, Stanford University; email: \texttt{ranxie@stanford.edu}.}\and
 Iain M. Johnstone\thanks{Department of Statistics, Stanford University; email: \texttt{imj@stanford.edu}.}
}

\begin{document}

\maketitle
\begin{abstract}
    We develop a nonparametric spectral estimation method for variance component covariance matrices in high-dimensional multivariate mixed effects models. Although classical MANOVA estimators are unbiased entrywise, their eigenvalues can be severely dispersed and biased when the number of traits is comparable to the number of realizations of the random effects. Building on a system of self-consistent equations that characterizes the limiting spectral distribution of a MANOVA estimator, we construct a kernel-based nonlinear transformation of the sample eigenvalues that targets the optimal rotation-invariant estimator under Frobenius loss. The resulting procedure corrects for high-dimensional eigenvalue spreading without imposing a parametric model on the unknown covariance spectrum. Our work is motivated by applications in quantitative genetics, where estimating genetic covariance components across many phenotypic traits is of central interest. We implement the method in a new Python package \texttt{mlmm-shrinkage} and compare the proposed estimator with classical estimators across several common experimental designs in this domain through simulations and a real-data application.
\end{abstract}

\section{Introduction}
We consider the $k$-level mixed linear model
\begin{align}\label{eq:general_model}
    Y = X\beta + \sum_{r=1}^k U_r \alpha_r, \quad \alpha_r \sim N(0, \Id _{I_r} \otimes \Sigma_r),
\end{align}
where $Y \in \mathbb{R}^{n \times p}$ is the observed trait matrix, $X \in \mathbb{R}^{n \times I_0}$ is the fixed effects design matrix, $\beta \in \mathbb{R}^{I_0 \times p}$ is the fixed effect, $U_r \in \mathbb{R}^{n \times I_r}$ are known design matrices, and $\alpha_r \in \mathbb{R}^{I_r \times p}$ are random effects. Our objective is to estimate $\Sigma_1,\ldots,\Sigma_k$ in a high-dimensional regime where $p,n,I_1,\ldots,I_k\to\infty$ proportionally.

Our estimators are built from quadratic form matrices
\begin{align}\label{eq:sos}
    S_n = \frac{1}{p} Y^T B Y, \quad \text{with } BX = 0,
\end{align}
where $B$ is known. This class includes sample covariance matrix, MANOVA-type estimators and MINQUE-type estimators. Our earlier work \cite{xie2024pargen} developed a parametric method-of-moments estimator with consistency and asymptotic normality guarantees, but it requires a parametric model for the eigenvalue distributions of $\Sigma_r$, which may be misspecified. In this paper, we develop a nonparametric shrinkage approach that preserves eigenvectors of a baseline estimator and shrinks eigenvalues toward a deterministic-equivalent oracle that minimizes the distance to the population covariance matrix $\Sigma_r$, followed by isotonic regression for additional stability. While this strategy is well understood in one-component Wishart setting \cite{ledoit2020analytical}, extending it to general $k$-component mixed models is nontrivial and constitutes our main contribution.

\subsection{Motivation from quantitative genetics}
In quantitative genetics, researchers measure multiple phenotypic traits for individuals and aim to infer the underlying sources of variation from these observable measurements. A classical approach is to decompose phenotypic variation into genetic and environmental components. Since the pioneering work of \cite{fisher1919xv}, linear mixed models have served as a fundamental framework for modeling such variance decompositions.

As a motivating example, consider the balanced full sibling design, where we are interested in the offspring of $I$ independent families. The $i$-th family has $J$ siblings, and the $j$-th sibling in the $i$-th family has $p$ measured phenotypic traits $Y_{ij}\in\mathbb{R}^p$. The observed trait is assumed to follow the model
\begin{align}\label{eq:fullsib.intro}
    Y_{ij} = \mu + \alpha_i + \epsilon_{ij}, \quad \alpha_i \sim N(0,\Sigma_1), \epsilon_{ij} \sim N(0,\Sigma_2),
\end{align}
where $\mu$ is the population mean, $\alpha_i$ is the family-level genetic effect, and $\epsilon_{ij}$ is the individual-level environmental effect. 

The genetic component is often further partitioned into additive and non-additive terms. The additive component is of primary biological interest, as it governs heritability and expected response to selection. Under the balanced full-sibling model, the additive genetic covariance matrix is approximately $2\Sigma_1$ \cite{falconer1996introduction, searle2009variance}. 

\begin{figure}
    \centering
    \includegraphics[width=0.7\linewidth]{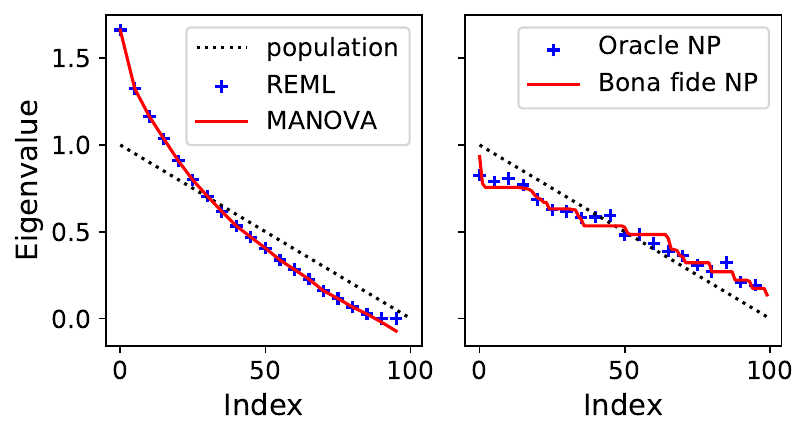}
    \caption{Comparison of oracle, bona fide and nonparametric shrinkage estimators for $\Sigma_1$ under the full sibling design.}
    \label{fig:intro}
\end{figure}

Figure \ref{fig:intro} previews the proposed nonparametric shrinkage estimator for $\Sigma_1$ in this setting. We compare the proposed estimator with two classical estimators commonly used in a low-dimensional setting with fixed $p$, Restricted Maximum Likelihood (REML) and Multivariate Analysis of Variance (MANOVA). As discussed in Section~\ref{sec:sim}, in the special case of a balanced full-sibling design, REML admits a closed-form solution \cite{Amemiya01051985}. The figure reveals substantial eigenvalue dispersion and bias for both REML and MANOVA. In addition, MANOVA yields negative eigenvalues, whereas REML is constrained to be positive semidefinite but exhibits a spurious null space. By contrast, the proposed shrinkage estimator more accurately tracks the target spectrum. Further details are provided in Section~\ref{sec:sim}.

\subsection{Related work}\label{subsec:prior.work}

Several approaches to estimating variance components in model~\eqref{eq:general_model} have been studied, and the present work draws on, and contrasts with, three strands of literature.

\paragraph{Classical approaches for mixed linear models with fixed $p$.} Restricted maximum likelihood (REML) is the default method in quantitative genetics for low-dimensional problems, with recent high-dimensional implementations under the full-sibling design developed by \cite{pavlyshyn2022}. However, REML is computationally demanding in high dimensions and currently lacks a comprehensive theory for spectral behavior, making it difficult to assess the quality of its eigenvalue estimates. MANOVA-type estimators of the form $Y^T B Y$ are unbiased and computationally tractable, but they may fail to be positive semidefinite and are therefore less used in practice. As we saw in Figure \ref{fig:intro}, both REML and MANOVA suffer from significant eigenvalue dispersion and bias in a high-dimensional setting.

\paragraph{Nonparametric shrinkage estimators for high-dimensional Wishart matrices.} For the Wishart setting ($k=1$), with $S_n=\frac{1}{p}\Sigma^{1/2}XX^T\Sigma^{1/2}$, with $\Sigma$ being a positive definite population covariance matrix, and $X\in\mathbb{R}^{p\times n}$ being a random matrix with i.i.d. entries, Ledoit and Peche \cite{Ledoit2011} showed that a deterministic-equivalent oracle that minimizes the distance to $\Sigma$ can be expressed as a function of the Stieltjes transform of the limiting spectral distribution of $S_n$.  Ledoit and Wolf \cite{ledoit2020analytical} subsequently proposed a bona fide nonparametric estimator that approximates this oracle via kernel smoothing of the empirical spectral distribution (ESD) of $S_n$. Oriol \cite{oriol2024asymptotic} extended the deterministic-equivalent oracle to separable covariance matrices ($k=1$), with  $S_n=\frac{1}{p}\Sigma^{1/2}XTX^{\ast}\Sigma^{1/2}$, with $\Sigma$ being a non-negative definite Hermitian matrix, and $T$ being a diagonal non-negative real matrix.

\paragraph{Random matrix theory for MANOVA-type matrices.} Fan and Johnstone \cite{fan2017eigenvalue} established a deterministic-equivalent characterization of the ESD of $S_n$ for general $k$-level models using operator-valued free probability theory in the high-dimensional setting. In particular, they proved that deterministic equivalent Stieltjes transform of the ESD of $S_n$, denoted $\tilde{m}_n$, is uniquely determined through a system of $2k$ equations involving component-specific quantities denoted by $\{\tilde{a}_n^{(r)},\tilde{b}_n^{(r)}\}_{r=1}^k$. Our earlier work \cite{xie2024clt} further established a central limit theorem for the linear spectral statistics of $S_n$.

\subsection{Approach and contributions}\label{subsec:contributions}

Our approach exploits a structural insight: the spectral information needed to construct the oracle estimator is encoded not only in the ESD of $S_n$ but also in $2k$ \textit{reweighted spectral measures}, defined through component-specific weights. We distinguish two families:
\begin{enumerate}
    \item The covariance-weighted measures $\mu_{n,a}^{(r)}$, whose Stieltjes transforms are 
    $$
    m_{n,a}^{(r)}(z) = p^{-1}\Tr((S_n - z\Id)^{-1}\Sigma_r),
    $$
    with weights determined by the covariance matrices $\Sigma_r$.
    \item The design-weighted measures $\mu_{n,b}^{(r)}$, whose Stieltjes transforms are 
    $$
    m_{n,b}^{(r)}(z) = p^{-1}\Tr((S_n - z\Id)^{-1}S_n^{(r)}),
    $$
    with weights determined by the known design matrices through $S_n^{(r)} = p^{-1}Y^TBU_rU_r^TBY$.
\end{enumerate}
In the one-component Wishart case studied by \cite{Ledoit2011}, the design-weighted family reduces to $p^{-1}\Tr((S_n - z\Id)^{-1}S_n) = 1 + zm_n(z)$, a function of the Stieltjes transform of the ESD alone. Consequently, this second family carries no additional information, and there is no need to estimate it separately. For $k > 1$, however, these design-weighted functionals are non-redundant and play an essential role.

The main contributions of this paper are as follows. First, we generalize the deterministic-equivalent oracle estimator developed in \cite{Ledoit2011,oriol2024asymptotic} from the one-component setting ($k=1$) to the general $k$-level setting. Building on the theoretical results of \cite{fan2017eigenvalue}, we derive the deterministic-equivalent oracle as a function of $\tilde{m}_n$, the deterministic equivalent Stieltjes transform of $S_n$, and the component-specific quantities $\{\tilde{a}_n^{(r)},\tilde{b}_n^{(r)}\}_{r=1}^k$. 

Second, we develop a nonparametric shrinkage estimator that approximates
the deterministic-equivalent oracle from observed data alone. This
requires estimating the deterministic-equivalent Stieltjes transform
$\tilde{m}_n$ and the component-specific quantities
$\{\tilde{a}_n^{(r)}, \tilde{b}_n^{(r)}\}_{r=1}^k$. In the
one-component setting, the oracle depends only on the Stieltjes
transform of the ESD, and it suffices to approximate $\tilde{m}_n$
via kernel smoothing \cite{ledoit2020analytical}. For general $k$,
estimating $\{\tilde{a}_n^{(r)}, \tilde{b}_n^{(r)}\}_{r=1}^k$
presents a new challenge. Building on \cite{fan2017eigenvalue}, we
show that these quantities are themselves deterministic equivalents of
the Stieltjes transforms of the covariance-weighted and
design-weighted measures $\mu_{n,a}^{(r)}$ and $\mu_{n,b}^{(r)}$,
respectively. This result may be of independent interest in random matrix theory. Because the design matrices $U_r$ are known, we extend the kernel smoothing
approach of \cite{ledoit2020analytical} to the design-weighted
measures to estimate $\{\tilde{b}_n^{(r)}\}_{r=1}^k$, and then
recover $\{\tilde{a}_n^{(r)}\}_{r=1}^k$ by inverting the
self-consistent system of $2k$ equations that links the two families.

Finally, we demonstrate the performance of the proposed estimator
through simulations under the full-sibling and full-sib half-sib
designs, as well as a real-data application. We implement our method in a new \textsf{Python} package {\texttt{mlmm-shrinkage}}, available at \url{https://github.com/ran-xie/mlmm-shrinkage}.


\subsection{Outline}

The remainder of the paper is organized as follows. Section~\ref{sec:algorithm} presents the nonparametric shrinkage estimator, proceeding from the oracle through its deterministic equivalent to the bona fide kernel-smoothed implementation. Section~\ref{sec:sim} reports simulation results and a real-data application under the full-sibling design. Section~\ref{sec:ar.br.borel} contains the proof of the main theoretical result. Technical details and supplementary results are deferred to the Appendix.

\section{The nonparametric shrinkage algorithm}\label{sec:algorithm}
We construct a nonparametric shrinkage estimator based on quadratic-form matrices $S_n=p^{-1} Y^TBY$ defined in \eqref{eq:sos}. For clarity of exposition, we first treat the full-rank case. The singular case is discussed in Section \ref{subsec:singular}. We focus on estimating $\Sigma_1$, since the same procedure applies to $\Sigma_r$ ($r=2,\ldots,k$) after choosing $B$ to target the corresponding variance component.

Our approach proceeds in three stages: (i) derive the oracle estimator minimizing Frobenius loss under knowledge of the true covariance, (ii) characterize its deterministic equivalent using free probability and random matrix theory, and (iii) implement a bona fide kernel-smoothed estimator approximating this deterministic equivalent from observed data.

Throughout the paper, we work under the following assumptions:
\begin{assum}\label{assum:minimum}
 There exist some constants $C,c>0$ such that
\begin{enumerate}
\item  $n,p,I_1,\ldots,I_k\rightarrow\infty$ such that $c<p/n<C$, $c<I_r/n<C$ for each $r=1\ldots,k$, 
    \item $\lVert B\rVert<C$, $\lVert \Sigma_r\rVert<C$, and $\lVert U_r\rVert<C$ for each $r=1,\ldots,k$.
    \item $\Sigma_r\neq0$, $U_r^TBU_r\neq 0$ for each $r=1,\ldots,k$.
 \end{enumerate}
\end{assum}
Let the spectral decomposition of $S_n$ be
\begin{align}
    S_n= \mathbb{U}_n\Lambda_n \mathbb{U}_n^T, \quad\Lambda_n=\text{diag}(\lambda_{n,1},\ldots,\lambda_{n,p}), \quad \mathbb{U}_n=(u_{n,1},\ldots,u_{n,p}),
\end{align}
where $\lambda_{n,1}\geq \ldots\geq \lambda_{n,p}$. Denote the ESD of $S_n$ by $\mu_n$, and its Stieltjes transform by $m_{n}(z)$, with
\begin{align}
    \mu_n=\frac{1}{p}\sum_{i=1}^p\delta_{\lambda_{n,i}},\quad m_{n}(z)=\frac{1}{p}\Tr \left[(S_n-z\Id )^{-1}\right].\label{eq:mu_n_m_n}
\end{align}  
Following \cite{fan2017eigenvalue}, the ESD of $S_n=p^{-1} Y^T B Y$ admits a deterministic equivalent characterized as follows.\footnote{Our assumption $\lVert B\rVert<C$ differs from the $n\lVert B\rVert<C$ condition in \cite{fan2017eigenvalue}, which accounts for the $\frac{1}{p}$ factor in $F$.} Let $I_+=\sum_{r=1}^kI_r$ and define
\begin{align}
    U=[\sqrt{I_1}U_1|\ldots|\sqrt{I_k}U_k],\quad F=\frac{1}{p}U^TBU.\label{eq:f}
\end{align}
Denote the spectral decomposition of $F$ as $F=RLR^T$, where $L\in\mathbb{R}^{l_0\times l_0}$ is full rank and diagonal, and $ R^T=[R_1^T,\ldots,R_k^T]$. Further define
\begin{align*}
   \tilde{R}^T=[\tilde{R}_1^T,\ldots,\tilde{R}_k^T], \quad \tilde{R}_1=\sqrt{p/I_r}R_r\in\mathbb{R}^{I_r\times l_0}.
\end{align*}
Then we have the following corollary of Theorem 1.2 in \cite{fan2017eigenvalue}.
\begin{lem}\label{lem:deteqv}
    Under Assumption \ref{assum:minimum}, for each $z\in\mathbb{C}^+$, there exists unique z-dependent values $\tilde{a}_n^{(1)},\ldots,\tilde{a}_n^{(k)}\in\mathbb{C}^+\cup\{0\}$, ${b}_n^{(1)},\ldots,{b}_n^{(k)}\in\overline{\mathbb{C}^+}$, and $\tilde{m}_n\in\mathbb{C}^+$, such that 
    \begin{align}
        \begin{cases}
        \tilde{m}_n&=-\frac{1}{p}\Tr ((z\Id +\sum_{r=1}^kb_n^{(r)}\Sigma_r)^{-1})\\
            \tilde{a}_n^{(r)}&=-\frac{1}{p}\Tr ((z\Id +\sum_{s=1}^kb_n^{(s)}\Sigma_s)^{-1}\Sigma_r),\quad r=1,\ldots,k\\
            {b}_n^{(r)}&=-\frac{1}{p}\Tr ((L^{-1}+\sum_{s=1}^k\tilde{a}_n^{(s)}\tilde{R}_s^T\tilde{R}_s)^{-1}\tilde{R}_r^T\tilde{R}_r),\quad r=1,\ldots,k. 
        \end{cases}\label{eq:b.n.r} 
    \end{align}
The function $\tilde{m}_n:\mathbb{C}^+\rightarrow\mathbb{C}^+$ defines the Stieltjes transform of a probability measure $\tilde{\mu}_n$ on $\mathbb{R}$ such that
\begin{align*}
    \tilde{m}_n(z)-m_n(z)\rightarrow_{a.s} 0,
\end{align*}
and $\tilde{\mu}_n-\mu_n\rightarrow0$ weakly almost surely.
\end{lem}
\begin{proof}
    The proof is given in Supplementary appendix \ref{sec:proofs}.
\end{proof}

\subsection{Reweighted spectral measures}
Define
\begin{align*}
    S_n^{(r)} = \frac{1}{p}Y^TBU_rU_r^TBY,\qquad r=1,\ldots,k.
\end{align*}
In addition to $\mu_n$, we introduce $2k$ reweighted spectral measures that encode component-specific information:
\begin{align}
\mu_{n,a}^{(r)}&=\frac{1}{p}\sum_{i=1}^pd_{n,i}^{(r)}\delta_{\lambda_{n,i}},\quad d_{n,i}^{(r)}=u_{n,i}^T\Sigma_ru_{n,i},\label{eq:mu.n.a}\\
    \mu_{n,b}^{(r)}&=\frac{1}{p}\sum_{i=1}^pt_{n,i}^{(r)}\delta_{\lambda_{n,i}},\quad t_{n,i}^{(r)}=u_{n,i}^TS_n^{(r)}u_{n,i},\label{eq:mu.n.b}
\end{align}
with Stieltjes transforms
\begin{align}
m_{n,a}^{(r)}&=\frac{1}{p}\sum_{i=1}^p\frac{d_{n,i}^{(r)}}{\lambda_{n,i}-z}=\frac{1}{p}\Tr ((\Lambda_n-z\Id )^{-1}\mathbb{U}_n^T\Sigma_r\mathbb{U}_n)= \frac{1}{p}\Tr ((S_n-z\Id )^{-1}\Sigma_r),\label{eq:m.n.a}\\
m_{n,b}^{(r)}&=\frac{1}{p}\sum_{i=1}^p\frac{t_{n,i}^{(r)}}{\lambda_{n,i}-z}=
    \frac{1}{p}\Tr ((\Lambda_n-z\Id )^{-1}\mathbb{U}_n^TS_n^{(r)}\mathbb{U}_n)=\frac{1}{p}\Tr ((S_n-z\Id )^{-1}S_n^{(r)}).\label{eq:m.n.b}
\end{align}

The family $\{m_{n,a}^{(r)}\}$ extends the functional $\frac{1}{p}\Tr ((S_n-z\Id )^{-1}\Sigma)$ identified by Ledoit and Peche in the one-component Wishart setting \cite{Ledoit2011}. The additional family $\{m_{n,b}^{(r)}\}$ is essential in multi-component models: in the one-component Wishart case,
\begin{align*}
    \frac{1}{p}\Tr ((S_n-z\Id )^{-1}S_n)=1+zm_n(z),
\end{align*}
so no new information is gained, whereas for $k>1$ these functionals are non-redundant.
\begin{lem}\label{lem:inversion.borel}
   Under Assumption \ref{assum:minimum}, for each $r=1,\ldots,k$,
    \begin{enumerate}
    \item The function $\tilde{a}_n^{(r)}(z)$ defined in Lemma \ref{lem:deteqv} maps the positive half complex plane to itself. It defines the Stieltjes transform of a positive Borel measure $\tilde{\mu}_{n,a}^{(r)}$ on $\mathbb{R}$ with $\tilde{\mu}_{n,a}^{(r)}(\mathbb{R})=\Tr (\Sigma_r)/p$ such that for each $z\in\mathbb{C}^+$,
        \begin{align*}
            \tilde{a}_n^{(r)}(z)-m_{n,a}^{(r)}(z)\rightarrow_{a.s.}0,
        \end{align*}
        and $\tilde{\mu}_{n,a}^{(r)}-\mu_{n,a}^{(r)}\rightarrow0$ weakly almost surely.
        \item Define 
\begin{align}\label{eq:tilde.b.n.r}
    \tilde{b}_n^{(r)}(z)=b_n^{(r)}(z)+\Tr (\tilde{R}_r^T\tilde{R}_rL)/p,
\end{align}
where $b_n^{(r)}$ is defined in Lemma \ref{lem:deteqv}.
The function $\tilde{b}_n^{(r)}(z)$ maps the positive half complex plane to itself. It
        defines the Stieltjes transform of a positive Borel measure $\tilde{\mu}_{n,b}^{(r)}$ on $\mathbb{R}$ with $\tilde{\mu}_{n,b}^{(r)}(\mathbb{R})=\frac{1}{p}\Tr (\Sigma_r)\sum_{s=1}^k\frac{1}{p}\Tr \left(\tilde{R}_rL\tilde{R}_s^T\tilde{R}_sL\tilde{R}_r^T\right)$ such that for each $z\in\mathbb{C}^+$,
        \begin{align*}
            \tilde{b}_n^{(r)}(z)-m_{n,b}^{(r)}(z)\rightarrow_{a.s.}0,
        \end{align*}
         and $\tilde{\mu}_{n,b}^{(r)}-\mu_{n,b}^{(r)}\rightarrow0$ weakly almost surely. 
    \end{enumerate}
\end{lem}
\begin{proof}
    The proof is provided in Section \ref{sec:ar.br.borel}.
\end{proof}
The proof builds on the operator-valued free probability framework developed in \cite{fan2017eigenvalue}. Part~(i) follows largely as an implicit consequence of their argument, although it is not stated there in the present form. Part~(ii) is substantially more delicate and does not follow from existing results. The obstacle is that $m_{n,b}^{(r)}(z)=p^{-1}\Tr ((S_n-z\Id )^{-1}S_n^{(r)})$ involves the auxiliary random matrix $S_n^{(r)}$, whereas the route in \cite{fan2017eigenvalue} treats functionals depending only on $S_n$. We overcome this by introducing an auxiliary matrix $\mathcal S_n$ and rewriting each $m_{n,b}^{(r)}$ as a resolvent trace involving $\mathcal S_n$. Because 
$\mathcal S_n$ is in general non-Hermitian, we need to generalize the arguments in \cite{fan2017eigenvalue} to handle this case. 

Set $\mathbb{R}^{\ast} := \mathbb{R}\setminus\{0\}$. If the Stieltjes 
transforms $\tilde m_n$ and $\{\tilde a_n^{(r)}, \tilde b_n^{(r)}\}_{r=1}^{k}$ 
admit boundary values on $\mathbb{R}^{\ast}$, then we can further show that the measures $\tilde\mu_n$, 
$\tilde\mu_{n,a}^{(r)}$, and $\tilde\mu_{n,b}^{(r)}$ are continuous  on $\mathbb{R}^{\ast}$.
\begin{assum}[Boundary condition]\label{assum:boundary}
    For all $x\in\mathbb{R}$, $x\neq 0$, and each $r=1,\ldots,k$, the limits 
    \begin{align*}
        \check{m}_n(x):= \lim_{z\in\mathbb{C}^+\rightarrow x}\tilde{m}_n(z),\quad
        \check{a}_n^{(r)}(x):= \lim_{z\in\mathbb{C}^+\rightarrow x}\tilde{a}_n^{(r)}(z),\quad
        \check{b}_n^{(r)}(x):= \lim_{z\in\mathbb{C}^+\rightarrow x}\tilde{b}_n^{(r)}(z),
    \end{align*}
    exist.
\end{assum}
\begin{lem}\label{lem:den}
    Under Assumptions \ref{assum:minimum} and \ref{assum:boundary}, the measures $\tilde{\mu}_n$, $\tilde{\mu}_{n,a}^{(r)}$, $\tilde{\mu}_{n,b}^{(r)}$ are continuous on $\mathbb{R}^{\ast}$ with density given by
    \begin{align*}
        d\tilde{F}_n(x)=\frac{1}{\pi}\Im [\check{m}_n(x)]dx,\quad d\tilde{F}_{n,a}^{(r)}(x)=\frac{1}{\pi}\Im [\check{a}_n^{(r)}(x)]dx,\quad d\tilde{F}_{n,b}^{(r)}(x)=\frac{1}{\pi}\Im [\check{b}_n^{(r)}(x)]d\tilde{F}_n(x).
    \end{align*}
\end{lem}
\begin{proof}
    The result follows from the Stieltjes inversion formula 
    \cite[Theorem~2.1]{silverstein1995analysis} together with a standard passage 
    to the limit \cite[Theorem~2.2]{silverstein1995analysis}.
\end{proof}

\subsection{Deterministic equivalent oracle estimator}\label{subsec:deteqv.oracle}

We evaluate performance under the squared Frobenius loss. For a $p\times p$ covariance matrix $\Sigma$ and an estimator $\hat{\Sigma}_n$, the squared Frobenius loss is defined as
\begin{align}
    \mathcal{L}_n^{\text{FR}}(\hat{\Sigma}_n,\Sigma)=\frac{1}{p}\Tr [(\hat{\Sigma}_n-\Sigma)^2].
\end{align}
Denote the eigenvalues of $\Sigma_1$ by
\begin{align*}
    \boldsymbol{\tau}_n = (\tau_{n,1},\ldots,\tau_{n,p}),\quad\tau_{n,1}\geq \ldots\geq \tau_{n,p}.
\end{align*}
We consider estimators that preserve the eigenvectors of $S_n$, and shrink the eigenvalues, of the form 
   $ \hat{\Sigma}_{n}=\mathbb{U}_nD_n \mathbb{U}_n^T$,
where $\mathbb{U}_n$ is the eigenvector matrix of $S_n$, and $D_n$ is a $p\times p$ diagonal matrix. A straightforward calculation shows that the oracle estimator minimizing $\mathcal{L}_n^{\text{FR}}$ over this class takes the form
\begin{align}\label{eq:oracle}
    \hat{\Sigma}_{n}^{\text{or}}=\mathbb{U}_nD_n^{\text{or}} \mathbb{U}_n^T, \quad D_n^{\text{or}}=\text{diag}(d_{n,1}^{\text{or}},\ldots,d_{n,p}^{\text{or}}), \quad d_{n,i}^{\text{or}}=u_{n,i}^T\Sigma_1u_{n,i}=d_{n,i}^{(1)},
\end{align}
where $d_{n,i}^{(1)}$ are the weights in the reweighted spectral measure $\mu_{n,a}^{(1)}$ defined in \eqref{eq:mu.n.a}. This oracle estimator is random, depending on the eigenvectors $\{u_{n,i}\}$ of $S_n$, and infeasible, as it requires knowledge of $\Sigma_1$. Generalizing the approach in \cite{Ledoit2011,oriol2024asymptotic} developed for Wishart matrices and separable covariance matrices corresponding to $k=1$, we now derive its deterministic equivalent, computed from the deterministic equivalent Stieltjes transform $\tilde{m}_n$ of the ESD of $S_n$ and $\tilde{a}_n^{(r)}$, $r=1,\ldots,k$, characterized in Lemma \ref{lem:deteqv}.

Denote the CDF of the ESD of $S_n$ by $F_n$, and the corresponding CDFs of the measures $\mu_{n,a}^{(r)}$ defined in (\ref{eq:mu.n.a}) by $F_{n,a}^{(r)}$, respectively. Denote the corresponding CDFs of the measures $\tilde\mu_n$ and $\tilde\mu_{n,a}^{(r)}$ defined in Lemma \ref{lem:deteqv} and Lemma \ref{lem:inversion.borel} by $\tilde{F}_n$ and $\tilde{F}_{n,a}^{(r)}$, respectively. When all the sample eigenvalues are distinct, it is straightforward to recover each $d_{n,i}^{(r)}$ as
\begin{align*}
    d_{n,i}^{(r)}=\lim_{\epsilon\rightarrow0^+}\frac{F_{n,a}^{(r)}(\lambda_i+\epsilon)-F_{n,a}^{(r)}(\lambda_i-\epsilon)}{F_n(\lambda_i+\epsilon)-F_n(\lambda_i-\epsilon)}.
\end{align*}
Under the boundary conditions given in Assumption \ref{assum:boundary}, by Lemma \ref{lem:den}, we have that
\begin{align*}
   d\tilde{F}_{n,a}^{(r)}(x)=\frac{\Im [\check{a}_n^{(r)}(x)]}{\Im [\check{m}_n(x)]}d\tilde{F}_n(x). 
 \end{align*}
For each eigenvalue $\lambda_{n,i}$, define the deterministic equivalent oracle estimator of $\tau_{n,i}$ as
\begin{align}\label{eq:deteqv.oracle}
    \check{d}_{n,i}=\frac{\Im [\check{a}_n^{(1)}(\lambda_{n,i})]}{\Im [\check{m}_n(\lambda_{n,i})]}.
\end{align}

To estimate eigenvalues of $\Sigma_r$ for $r \neq 1$, one replaces $\check{a}_n^{(1)}$ with $\check{a}_n^{(r)}$ in the definition above, after appropriately designing the projection matrix $B$ to target that covariance component.

\subsection{Bona fide nonparametric estimator}
The construction of our nonparametric shrinkage estimator, which we refer to as the \textit{bona fide} estimator, involves several numerical steps. We first approximate the deterministic equivalent oracle using kernel smoothing techniques and then refine the resulting estimates via isotonic regression. The complete procedure is summarized in Algorithm \ref{alg:full}. 

The key task is to approximate the denominator $\Im[\check{m}_n]$ and the numerator $\Im[\check{a}_n^{(1)}]$. In \cite{ledoit2020analytical}, the authors propose estimating the former by smoothing the ESD $\mu_n$ of $S_n$ to obtain a continuous density $\hat f_n$, and then using its Stieltjes transform $\hat m_n$ as an approximation to $\check m_n$. This strategy remains valid for the denominator but does not extend directly to the numerator. In the Wishart setting of \cite{ledoit2020analytical}, $a_n$ is a function of $m_n$, so estimating $m_n$ is sufficient. With multiple variance components, this functional reduction no longer holds. Estimating $a_n^{(1)}$ therefore requires a new approach that explicitly uses the multi-level variance structure. We first adapt their method to estimate $\check m_n$, and then present our estimator of $\check a_n^{(1)}$.

We adopt the kernel assumptions of \cite{ledoit2020analytical}.
\begin{assum}[Assumption 4.1 in \cite{ledoit2020analytical}]\label{assum:kernel}
Let $k(x)$ denote a continuous, symmetric, nonnegative probability density function (p.d.f.) whose support is a compact interval $[-R,R]$, with mean zero
and variance one. We assume throughout that this kernel satisfies the following conditions:
\begin{enumerate}
    \item Its Hilbert transform $\mathcal{H}_k$ exists and is continuous on $\mathbb{R}$.
    \item Both the kernel $k$ and its Hilbert transform $\mathcal{H}_k$ are functions of bounded variation.
\end{enumerate}
\end{assum}
\begin{definition}
    The Hilbert transform of a real function $f$ is defined as
    \begin{align*}
        \mathcal{H}_f(x)=\frac{1}{\pi}\text{PV}\int_{-\infty}^\infty\frac{f(t)}{t-x}dt,
    \end{align*}
where $\text{PV}$ denotes the Cauchy principal value
\begin{align*}
    \text{PV}\int_{-\infty}^\infty\frac{f(t)}{t-x}dt=\lim_{\epsilon\rightarrow 0^+}\left[\int_{-\infty}^{x-\epsilon}\frac{f(t)}{t-x}dt+\int_{x+\epsilon}^\infty\frac{f(t)}{t-x}dt\right].
\end{align*}
\end{definition}
Following \cite{ledoit2020analytical}, we use a proportional bandwidth to account for the phenomenon that in a proportional regime, large eigenvalues get spread out more than small ones. For a detailed discussion of this choice, see \cite{ledoit2020analytical}. For each $\lambda_{n,i}\neq 0$, define
\begin{align*}
    h_{n,i}=h_n\max\{|\lambda_{n,i}|,\epsilon\}.
\end{align*}
The distinction from their setting is that, in our model, the support of the ESD of $S_n$ may include the origin. In contrast, the eigenvalues in their setting are positive and bounded away from zero. Their proportional bandwidth is recovered by taking $\epsilon=0$. Section \ref{sec:isotonic} describes our choice of $h_n$ and $\epsilon$.

A key identity used in \cite{ledoit2020analytical} is the boundary value formula for the Stieltjes transform. For kernels satisfying Assumption \ref{assum:kernel}, let $m_k$ denote its Stieltjes transform, then for any $x\neq 0$,
\begin{align*}
    \lim_{\eta\rightarrow 0^+}m_k(x+i\eta)= \lim_{\eta\rightarrow 0^+}\int\frac{k(t)}{t-(x+i\eta)}\,dt
    =\pi\mathcal{H}_{k}(x)+i\pi k(x).
\end{align*}
Motivated by this identity, let $\hat f_n$ be the kernel estimator of the ESD $\mu_n$, and let $\mathcal{H}_{\hat f_n}$ denote its Hilbert transform:
\begin{align*}
    \hat{f}_n(x)
    = \frac{1}{p}\sum_{i=1}^p\frac{1}{h_{n,i}}k\!\left(\frac{x-\lambda_{n,i}}{h_{n,i}}\right),\quad
    \mathcal{H}_{\hat{f}_n}(x)
    = \frac{1}{p}\sum_{i=1}^p\frac{1}{h_{n,i}}\mathcal{H}_k\!\left(\frac{x-\lambda_{n,i}}{h_{n,i}}\right).
\end{align*}
We then approximate $\check m_n$ by
\begin{align*}
\hat{m}_{n}(x)
:=\lim_{\eta\rightarrow 0^+}\int\frac{\hat{f}_n(t)}{t-(x+i\eta)}\,dt
=\pi\mathcal{H}_{\hat{f}_n}(x)+i\pi \hat{f}_n(x).
\end{align*}

\begin{algorithm}[!htbp]
    \caption{Nonparametric Shrinkage Estimation for $\Sigma_1$ ($S_n$ full rank)}\label{alg:full}
    \textbf{Input:}
    \begin{algorithmic} 
    \State Number of observations $n$, number of eigenvalues $p$, number of levels $k$, number of observations per level $I_1,\ldots,I_k$, global bandwidth $h_n=n^{-1/3}$
    \State Observations $Y\in\mathbb{R}^{n\times p}$, design matrices $(X, \{U_r\}_{r = 1}^k)$, symmetric matrix $B$
    \end{algorithmic}
    \textbf{Procedure:}
    \begin{algorithmic}[1] 
    \State Compute some useful quantities: $U=[\sqrt{I_1}U_1|\ldots|\sqrt{I_k}U_k]$, $F=\frac{1}{p}U^TBU$ with spectral decomposition $F=RLR^T$, where $L\in\mathbb{R}^{l_0\times l_0}$ is full rank and diagonal for some rank $l_0$, and $R^T=[R_1^T,\ldots,R_k^T]$ with each $R_r\in\mathbb{R}^{I_r\times l_0}$. $\tilde{R}_r=(p/I_r)R_r$.
    \State Construct base estimator $S_n=\frac{1}{p}Y^TBY$ with spectral decomposition $ S_n= \mathbb{U}_n\Lambda_n \mathbb{U}_n^T$, where $\Lambda_n=\text{diag}(\lambda_{n,1},\ldots,\lambda_{n,p})$, $\lambda_{n,1}\geq \ldots\geq \lambda_{n,p}$,  $\mathbb{U}_n=(u_{n,1},\ldots,u_{n,p})$. 
    \State Compute proportional bandwidths $h_{n,i}=h_n\lambda_{n,i}.$
    \State Compute the kernel smoothed p.d.f of the ESD of $S_n$ and its Hilbert transform as
    \[
    \hat{f}_n(x) = \tfrac{1}{p} {\textstyle\sum}_{i=1}^p \tfrac{1}{h_{n,i}} \kappa^E \Bigl( \tfrac{x-\lambda_{n,i}}{h_{n,i}} \Bigr), \quad 
    \mathcal{H}_{\hat{f}_n}(x) = \tfrac{1}{p} {\textstyle\sum}_{i=1}^p \tfrac{1}{h_{n,i}} \mathcal{H}_{\kappa^E} \Bigl( \tfrac{x-\lambda_{n,i}}{h_{n,i}} \Bigr),
\]
     where $\kappa^E$ and $\mathcal{H}_{\kappa^E}$ are respectively the Epanechnikov kernel and its Hilbert transform as in (\ref{eq:kappa.E}). 
     \State Obtain the kernel smoothed Stiletjes transform
     $\hat{m}_{n}(x)=\pi\mathcal{H}_{\hat{f}_n}(x)+i\pi \hat{f}_n(x).$
    \For{\texttt{$r \gets 1$ to $k$}}
    \State Construct matrix $S_n^{(r)} = \frac{1}{p}Y^TBU_rU_r^TBY$, and compute weights $t_{n,i}^{(r)}=u_{n,i}^TS_n^{(r)}u_{n,i}$.
    \State Compute the weighted measure and its Hilbert transform   
    \[
        \hat{g}_n^{(r)}(x) = \tfrac{1}{p} {\textstyle\sum}_{i=1}^p \tfrac{t_{n,i}^{(r)}}{h_{n,i}} \kappa^E \Bigl( \tfrac{x-\lambda_{n,i}}{h_{n,i}} \Bigr), \quad 
        \mathcal{H}_{\hat{g}_n^{(r)}}(x) = \tfrac{1}{p} {\textstyle\sum}_{i=1}^p \tfrac{t_{n,i}^{(r)}}{h_{n,i}} \mathcal{H}_{\kappa^E} \Bigl( \tfrac{x-\lambda_{n,i}}{h_{n,i}} \Bigr).
    \]
   \State Obtain the Stiletjes transform of the weighted measure
    $\hat{b}_{n}^{(r)}(x)=\pi\mathcal{H}_{\hat{g}_n^{(r)}}(x)+i\pi \hat{g}_n^{(r)}(x).$
    \EndFor
    \State For each $\lambda_{n,i}$, set $(\hat{a}_{n}^{(1)}(\lambda_{n,i}),\ldots,\hat{a}_{n}^{(k)}(\lambda_{n,i}))$ as the solution to the system of equations
    \begin{align*}
    \begin{cases}
            \hat{b}_{n}^{(r)}(\lambda_{n,i}) &= -\frac{1}{p}\Tr \left((L^{-1}+\sum_{s=1}^k\hat{a}_{n}^{(s)}(\lambda_{n,i})\tilde{R}_s^T\tilde{R}_s)^{-1}\tilde{R}_r^T\tilde{R}_r\right)+\frac{1}{p}\Tr (\tilde{R}_r^T\tilde{R}_rL).
    \end{cases}
\end{align*}
    \State Denote the shrinkage estimator for the $i$-th largest eigenvalue of $\Sigma_1$ by $\hat{d}_{n,i}$. Set
    $$
    \hat{d}_n=(\hat{d}_{n,1},\ldots,\hat{d}_{n,p}),\quad\hat{d}_{n,i}=\frac{\Im [\hat{a}_n^{(1)}(\lambda_{n,i})]}{\Im [\hat{m}_n(\lambda_{n,i})]}.
    $$
   \State (optionally) Perform isotonic regression on $\hat{d}_n$ as described in Section \ref{sec:isotonic}.
    \end{algorithmic}
    \textbf{Output:}
    \begin{algorithmic}
     \State Nonparametric shrinkage estimator $\hat{\Sigma}_1=\mathbb{U}_n\text{diag}(\hat{d}_n)\mathbb{U}_n^T$
    \end{algorithmic}
\end{algorithm}

\subsubsection{Kernel-based approximation of $\check{a}_n^{(1)}$}
Estimating the numerator is substantially more challenging than estimating $\check m_n$ and is a main contribution of our framework. In the one-component Wishart model, the oracle can be written as a function of $m_n$ alone. With multiple variance components, this reduction is no longer available. Our construction instead exploits three structural facts: (i) the design matrices $U_1,\dots,U_k$ are known, (ii) Lemma \ref{lem:inversion.borel} identifies each $\tilde{b}_n^{(r)}$ as the deterministic equivalent to the functionals $m_{n,b}^{(r)}$ defined in \eqref{eq:m.n.b}, with $S_n^{(r)}$ computed from $U_r$ and the observed data $Y$, and (iii) Lemma \ref{lem:deteqv} establishes a one-to-one correspondence between $\{\tilde a_n^{(r)}\}_{r=1}^k$ and $\{\tilde b_n^{(r)}\}_{r=1}^k$ through a coupled fixed-point system, so that estimates of $\{\tilde b_n^{(r)}\}_{r=1}^k$ can be mapped to estimates of $\{\tilde a_n^{(r)}\}_{r=1}^k$.

Our procedure is to first estimate $\tilde b_n^{(r)}$ from data, since the design-dependent weights are observable, and then recover $\tilde a_n^{(r)}$ by inverting the system linking $\tilde a_n^{(r)}$ and $\tilde b_n^{(r)}$.

For each $r=1,\ldots,k$, let $\hat{g}_n^{(r)}$ be the kernel estimator of the empirical measure $\mu_{n,b}^{(r)}$, and let $\mathcal{H}_{\hat{g}_n^{(r)}}$ denote its Hilbert transform:
\begin{align*}
    \hat{g}_n^{(r)}(x) = \frac{1}{p}\sum_{i=1}^p\frac{t_{n,i}^{(r)}}{h_{n,i}}k\left(\frac{x-\lambda_{n,i}}{h_{n,i}}\right),\quad \mathcal{H}_{\hat{g}_n^{(r)}}(x) = \frac{1}{p}\sum_{i=1}^p\frac{t_{n,i}^{(r)}}{h_{n,i}}\mathcal{H}_k\left(\frac{x-\lambda_{n,i}}{h_{n,i}}\right).
\end{align*}
Then similarly as we estimated $\check{m}_n$, we  can estimate $\check{b}_n^{(r)}(x):= \lim_{\eta\rightarrow0^+}\tilde{b}_n^{(r)}(x+i\eta)$ by
\begin{align*}
 \hat{b}_{n}^{(r)}(x):=   \lim_{\eta\rightarrow 0^+}\int\frac{\hat{g}_n^{(r)}(t)}{t-(x+i\eta)}dt&=\pi\mathcal{H}_{\hat{g}_n^{(r)}}(x)+i\pi \hat{g}_n^{(r)}(x).
\end{align*}

Combining (\ref{eq:b.n.r}) and (\ref{eq:tilde.b.n.r}), for each $x\neq0$, set $(\hat{a}_{n}^{(1)}(x),\ldots,\hat{a}_{n}^{(k)}(x))$ as the solution to the system of equations
\begin{align}\label{eq:solve.bonafide}
\begin{cases}
    \hat{b}_{n}^{(r)} &= -\frac{1}{p}\Tr \left((L^{-1}+\sum_{s=1}^k\hat{a}_{n}^{(s)}\tilde{R}_s^T\tilde{R}_s)^{-1}\tilde{R}_r^T\tilde{R}_r\right)+\frac{1}{p}\Tr (\tilde{R}_r^T\tilde{R}_rL).
\end{cases}
\end{align}
While the above system is in general solved numerically, \cite[Appendix~A]{fan2017eigenvalue} derive simplified expressions that admit closed-form solutions in common experimental designs, including one-way classification, balanced nested classification and replicated crossed two-way classification.

Finally, for each $\lambda_{n,i}\neq 0$, we set the nonparametric estimate of $\tau_{n,i}$, the $i$-th largest eigenvalue of $\Sigma_1$, as
\begin{align*}
    \hat{d}_{n,i} &= \frac{\Im \hat{a}_n^{(1)}(\lambda_{n,i})}{\Im \hat{m}_n(\lambda_{n,i})}= \frac{\Im \hat{a}_n^{(1)}(\lambda_{n,i})}{\pi\hat{f}_n(\lambda_{n,i})}.
\end{align*}

\subsection{The singular case}\label{subsec:singular}
For Wishart matrices, \cite[Appendix C]{ledoit2020analytical} derives the deterministic equivalent oracle shrinkage formula applied to the zero eigenvalues of $S_n$, and propose a bona fide estimator. Their method extends the kernel-smoothing technique applied to the nonzero eigenvalues and involves only the Stieltjes transform of the ESD of $S_n$. For separable covariance matrices, \cite[Theorem 2.4]{oriol2024asymptotic} obtain an analogous deterministic equivalent oracle shrinkage formula, again expressed through the Stieltjes transform of the ESD of $S_n$ alone. With multiple components ($k>1$), such reduction is no longer available, and both the oracle and the bona fide estimators take a fundamentally different form, determined through the components-specific quantities $\{a_r^{(r)}(z),b_r^{(r)}(z)\}_{r=1}^k$. Supplementary Appendix~\ref{sec:singular.separable} verifies that, when specialized to $k=1$, our deterministic equivalent oracle shrinkage estimator recovers the formulae of
\cite{ledoit2020analytical} and \cite{oriol2024asymptotic}.

When $S_n$ is singular, its empirical spectral distribution places a non-negligible point mass at $0$ in addition to a continuous component on $\mathbb{R}^*$. For the nonzero eigenvalues of $S_n$, the derivations of the deterministic equivalent oracle estimator in Section \ref{subsec:deteqv.oracle} continue to apply on $\mathbb{R}^*$. Practically, we therefore estimate the continuous part of the ESD by kernel smoothing restricted to $\mathbb{R}^*$ and adopt a similar strategy to obtain shrinkage estimates for $\lambda_{n,i}\neq 0$. We defer the algorithmic details to Supplementary Appendix \ref{app:singular}, with a detailed algorithm given in Algorithm \ref{alg:singular}.

It remains to specify the treatment of the null space of $S_n$. Let
$\kappa_n = \mathrm{rank}(S_n)$, so the null space has dimension
$p - \kappa_n$. Although the eigenvectors spanning the null space are
not uniquely identified, their \emph{average} contribution is. For $r = 1,\ldots,k$, define the average
component-specific quantities
\begin{align*}
    d_{n,0}^{(r)}
    := \frac{1}{p-\kappa_n}\sum_{i=1}^{p}
    d_{n,i}^{(r)}\mathbbm{1}\{\lambda_{n,i}=0\},
    \qquad
    d_{n,0}^{\,\mathrm{or}} := d_{n,0}^{(1)} .
\end{align*}
Then $d_{n,0}^{\,\mathrm{or}}$ is the oracle shrinkage estimator of the average of the $p-\kappa_n$ smallest eigenvalues of $\Sigma_1$, denoted $\tau_{n,0}$, i.e.,
\begin{align*}
    \tau_{n,0}
    \;:=\; \frac{1}{p-\kappa_n}\sum_{i=1}^{p}
    \tau_{n,i}\mathbbm{1}\{\lambda_{n,i}=0\}.
\end{align*}
By definition of the reweighted spectral measure $\mu_{n,a}^{(1)}$ in (\ref{eq:mu.n.a}), we can alternatively express $d_{n,0}^{\text{or}}$ as
\begin{align*}
    d_{n,0}^{\text{or}} = \frac{p}{p-\kappa_n}\mu_{n,a}^{(1)}(0),
\end{align*}
which motivates the deterministic-equivalent counterparts
\begin{align*}
    \tilde d_{n,0}^{(r)}
    := \frac{p}{p-\kappa_n}\,\tilde\mu_{n,a}^{(r)}(\{0\}),
    \qquad
    \check d_{n,0}^{\,\mathrm{or}}
    := \tilde d_{n,0}^{(1)},
\end{align*}
where $\check{d}_{n,0}^{\,\mathrm{or}}$ is the deterministic equivalent oracle shrinkage estimator of $\tau_{n,0}$.
Leveraging the multiple component structure, we obtain an explicit characterization of
$\{\tilde d_{n,0}^{(r)}\}_{r=1}^{k}$ through the boundary values of the
Stieltjes transforms $\{\tilde b_n^{(r)}\}_{r=1}^{k}$ at the origin. Denote the restriction of $\tilde{\mu}_{n,a}^{(r)}$ to $\mathbb{R}^*$ by $\tilde{\mu}_{n,a,\mathbb{R}^*}^{(r)}$, with Stieltjes transform $\tilde{a}_{n,\mathbb{R}^*}^{(r)}$.
\begin{lem}\label{lem:singular}
    Under Assumptions \ref{assum:minimum} and \ref{assum:boundary}, and assuming that for each $r=1,\ldots,k$, $\Sigma_r$ is positive definite, and $\lim_{z\in\mathbb{C}^+\rightarrow0}z\tilde{a}_{n,\mathbb{R}^*}^{(r)}(z)=0$, then
    \begin{align*}
        \check{b}_n^{(r)}(0)&:= \lim_{z\in\mathbb{C}^+\rightarrow0}\tilde{b}_n^{(r)}(z)=\frac{1}{p}\Tr (\tilde{R}_r^T\tilde{R}_rL),\\
        \check{b}_n^{(r)}{}'(0)&:=\lim_{z\in\mathbb{C}^+\rightarrow0}\frac{\tilde{b}_n^{(r)}(z)-\check{b}_n^{(r)}(0)}{z}=\frac{1}{p-\kappa_n}\Tr (({\textstyle{\sum}}_{s=1}^k\check{d}_{n,0}^{(s)}\tilde{R}_s^T\tilde{R}_s)^{-1}\tilde{R}_r^T\tilde{R}_r).
    \end{align*}
\end{lem}
\begin{proof}
    The proof is provided in Supplementary Appendix \ref{app:proof.singular}.
\end{proof}
Let $m_{n,b,\mathbb{R}^*}^{(r)}$ denote the Stieltjes transform of the restriction of $\mu_{n,b}^{(r)}$ to $\mathbb{R}^*$,
\begin{align*}
    m_{n,b,\mathbb{R}^*}^{(r)}(z)=\frac{1}{\kappa_n}\sum_{i=1}^{p}\frac{t_{n,i}^{(r)}}{\lambda_{n,i}-z}\mathbbm{1}[\lambda_{n,i}\neq0].
\end{align*}
Then we estimate $\check{b}_n^{(r)}{}'(0)$ by
\begin{align}\label{eq:singular.b.n.r.prime}
    \hat{b}_{n}^{(r)}{}'(0):=\frac{\kappa_n}{p}m_{n,b,\mathbb{R}^*}^{(r)}{}'(0)=\frac{1}{p}\sum_{i=1}^{p}\frac{t_{n,i}^{(r)}}{\lambda_{n,i}^2}\mathbbm{1}[\lambda_{n,i}\neq0].
\end{align}
Finally, we set the nonparametric estimate of $\tau_{n,0}$ as $\hat{d}_{n,0}:=\hat{d}_{n,0}^{(1)}$, where $(\hat d_{n,0}^{(1)},\ldots,\hat d_{n,0}^{(k)})$ is the solution to the system
    \begin{align}\label{eq:singular.system}
    \begin{cases}
      \hat{b}_{n}^{(r)}{}'(0)=\frac{1}{p-\kappa_n}\Tr (({\textstyle{\sum}}_{s=1}^k\hat{d}_{n,0}^{(s)}\tilde{R}_s^T\tilde{R}_s)^{-1}\tilde{R}_r^T\tilde{R}_r).
    \end{cases}
\end{align}

\subsubsection{Isotonic regression and implementation details}\label{sec:isotonic}
In finite samples, the raw nonparametric estimates $\hat d_{n,1},\ldots,\hat d_{n,p}$ in the $k$-level model can exhibit substantial local irregularity, as illustrated by the simulation results in Section~\ref{sec:sim}. To enforce the structural ordering and reduce estimation noise, we apply isotonic regression as a post-estimation regularization step.

This choice is aligned with the covariance estimation literature. In the single-component case ($k=1$), Stein’s classical estimator, which preserves the sample eigenvectors while shrinking the corresponding eigenvalues \cite{stein1975estimation}, is often coupled with order-preserving adjustments such as isotonic regression; see, for example, Stein’s isotonizing algorithm described in the appendix of \cite{lin1985monte}. More recently, Ledoit and Wolf \cite{ledoit2020analytical} showed that their analytic nonparametric estimator for Wishart matrices is typically sufficiently regular in the $k=1$ setting, so that an additional isotonic step is not required. Our estimator reduces to theirs when $k=1$. However, for general $k$-level models, the raw sequence $\{\hat d_{n,i}\}_{i=1}^p$ is empirically more noisy, so an isotonising regularization step remains useful.

Accordingly, after computing $\hat d_{n,1},\ldots,\hat d_{n,p}$, we obtain the isotonic estimator via the Pool-Adjacent-Violators Algorithm (PAVA):
\begin{align}
    \hat{d}_n^{\text{iso}}=\arg\min_{\beta\in\mathbb{R}^p}\sum_{i=1}^p(\hat{d}_{n,i}-\beta_i)^2\text{ subject to }\beta_1\leq \ldots\leq \beta_p.
\end{align}
The algorithm proceeds as follows:
\begin{enumerate}
\item Initialize by setting $\beta_i = \hat{d}_{n,i}$ for $i = 1, \ldots, p$. Create $p$ blocks, where the $r$-th block contains the single index $r$.
\item If $\beta_1 \leq \cdots \leq \beta_p$, terminate and output $\hat{d}_n^{\text{iso}} = \beta$.
\item Otherwise, identify the first index $i$ such that $\beta_i > \beta_{i+1}$. Merge the blocks containing the indices $i$ and $i+1$, and reset the values of all elements in the merged block to their block average. Return to Step 2.
\end{enumerate}

For the numerical implementation, we use the Epanechnikov kernel and set the global bandwidth to $h_n=n^{-1/3}$, following \cite{ledoit2020analytical}. From Proposition 4.1 in \cite{ledoit2020analytical}, for $x\in\mathbb{R}$, the Epanechnikov kernel and its Hilbert transform are respectively

\begin{align}\label{eq:kappa.E}
    \kappa^E(x) = \frac{3}{4\sqrt{5}}\left[1-\frac
    {x^2}{5}\right]^+,\quad \mathcal{H}_{\kappa^E}(x)=\begin{cases}
        -\frac{3x}{10\pi}+\frac{3}{4\sqrt{5}\pi}
        \left(1-\frac{x^2}{5}\right)\log\left|\frac
        {\sqrt{5}-x}{\sqrt{5}+x}\right|\quad &\text
        {if }|x|\neq \sqrt{5}\\
        -\frac{3x}{10\pi}&\text{if }|x|= \sqrt{5}
    \end{cases},
\end{align}
where $\log$ denotes the natural logarithm.
To ensure stability near the origin, we set the individual bandwidths as $h_{n,i}=h_n\max\{|\lambda_{n,i}|,\epsilon\}$, where $\epsilon$ is a small tuning hyperparameter. $\epsilon$ controls the minimal bandwidth near the singularity at 0, balancing variance inflation due to undersmoothing against global bias from oversmoothing. Here we tune $\epsilon$ using the standard likelihood cross-validation (LCV) approach \cite{duin1976choice,HABBEMA1974}, finding 
\begin{align*}
    \epsilon_{\rm cv} = \arg\max_{\epsilon} \sum_{i=1}^p \log \hat{f}_{n,-i}(\lambda_i),
\end{align*}
where $\hat{f}_{n,-i}$ is the kernel estimate leaving out the $i$-th empirical eigenvalue $\lambda_i$.

\section{Experiments under the full sibling model}
We use the balanced full-sibling model in (\ref{eq:fullsib.intro}) as an illustrative case, as it yields closed-form expressions for $\tilde{a}_n^{(r)}(z)$ and $\tilde{b}_n^{(r)}(z)$, $r=1,2$.  In Supplementary Appendix \ref{app:halfsib}, we provide more involved simulations under the unbalanced full sib half sib model.

We first specialize the algorithm to this design, then report simulation and real-data results. In matrix form, (\ref{eq:fullsib.intro}) is given by
\begin{align*}
    Y=X\beta+U_1\alpha_1+U_2\alpha_2,
\end{align*}
where $X=\mathbbm{1}_n$ is the fixed effects design matrix, $\beta=\mu^T$ is the fixed effect, $U_1=\text{diag}(\mathbbm{1}_{J},\ldots,\mathbbm{1}_{J})\in\mathbb{R}^{n\times I}$ is the membership matrix that assigns
each sibling to the corresponding family, $U_2=I_n$ is the membership matrix that assigns each individual to the corresponding individual effect, and $\alpha_1\in\mathbb{R}^{I\times p}$, $\alpha_2\in\mathbb{R}^{n\times p}$ contain the stacked family and individual effects, respectively.

Define projection matrices 
\begin{align*}
\pi_1 = U_1(U_1^TU_1)^{-1}U_1^T&-\mathbbm{1}_n(\mathbbm{1}_n^T\mathbbm{1}_n)^{-1}\mathbbm{1}_n^T,\quad \pi_2=I_n-U_1(U_1^TU_1)^{-1}U_1^T.
\end{align*}
To estimate $\Sigma_1$, we first compute the unbiased MANOVA estimator of  $\Sigma_1$, which is given by
\begin{align*}
    S_n&=\frac{1}{p}Y^TBY\in\mathbb{R}^{p\times p},\quad B=\frac{p}{J(I-1)}\pi_1-\frac{p}{J(n-I)}\pi_2,
\end{align*}
with $\mathbb{E}(S_n)=\Sigma_1$. Let $F\in\mathbbm{R}^{(I+n)\times (I+n)}$ be defined as in \eqref{eq:f}, then 
\begin{align*}
F_{11}=\frac{I}{p}\tilde{R}_1 L\tilde{R}_1^T=\frac{I}{p}U_1^TBU_1=\frac{p}{J(I-1)}\frac{I}{p}U_1^T\pi_1U_1.
\end{align*}
\begin{align*}
F_{22}=R_2 LR_2^T=\frac{n}{p}\tilde{R}_2 L\tilde{R}_2^T=\frac{n}{p}U_2^TBU_2=\frac{n}{p}B.
\end{align*}
Since $\text{Tr}(U_1^T\pi_1U_1)=\text{Tr}(U_1^TU_1)-\frac{1}{n}\text{Tr}(U_1^T\mathbbm{1}_n\mathbbm{1}_n^TU_1)=n-J=J(I-1)$, we have
\begin{align}
\frac{1}{p}\text{Tr}(\tilde{R}_1^T\tilde{R}_1L)&=\frac{1}{J(I-1)}\text{Tr}(U_1^T\pi_1U_1)=1,\label{eq:full.sib.tr.r}\\
\frac{1}{p}\text{Tr}(\tilde{R}_2^T\tilde{R}_2L)&=\frac{1}{p}\text{Tr}B=0.\nonumber
\end{align}

By Corollary 2.1 of \cite{fan2017eigenvalue}, the system in Lemma \ref{lem:deteqv} simplifies to 
\begin{align*}
\begin{cases}
    \tilde{m}_n&=-\frac{1}{p}\text{Tr}((z\Id +b_1\Sigma_1+b_2\Sigma_2)^{-1}),\\
    \tilde{a}_n^{(1)}&=-\frac{1}{p}\text{Tr}((z\Id +b_1\Sigma_1+b_2\Sigma_2)^{-1}\Sigma_1),\\
    \tilde{a}_n^{(2)}&=-\frac{1}{p}\text{Tr}((z\Id +b_1\Sigma_1+b_2\Sigma_2)^{-1}\Sigma_2),\\
    b_n^{(1)}&=-(1+(p/I)\tilde{a}_n^{(1)}+(p/n)\tilde{a}_n^{(2)})^{-1},\\
    b_n^{(2)}&=J^{-1}(J-1)(J-1-(p/n)\tilde{a}_n^{(2)})^{-1}+J^{-1}{b_n^{(1)}}
\end{cases}
\end{align*}
By definition of $\tilde{b}_n^{(r)}$ given in \eqref{eq:tilde.b.n.r}, and the quantities $\frac{1}{p}\text{Tr}(\tilde{R}_r^T\tilde{R}_rL)$ obtained in \eqref{eq:full.sib.tr.r}, we have 
\begin{align*}
    \begin{cases}
        \tilde{b}_n^{(1)}&=-(1+(p/I)\tilde{a}_n^{(1)}+(p/n)\tilde{a}_n^{(2)})^{-1}+1,\\
    \tilde{b}_n^{(2)}&=J^{-1}(J-1)(J-1-(p/n)\tilde{a}_n^{(2)})^{-1}+J^{-1}(\tilde{b}_n^{(1)}-1).
    \end{cases}
\end{align*}
Inverting the last two equations yields closed-from solutions
\begin{align*}
    \hat{a}_n^{(2)}&=\frac{n}{p}(J-1)\left(1-\frac{1}{J\hat{b}_n^{(2)}-\hat{b}_n^{(1)}+1}\right),\\
    \hat{a}_n^{(1)}&=-\frac{I}{p}\left(1+\frac{p}{n}\hat{a}_n^{(2)}+\frac{1}{\hat{b}_n^{(1)}-1}\right).
\end{align*}

\subsection{Simulation study}\label{sec:sim}
We consider a balanced full-sibling design with $I={500}$ families, each containing $J={4}$ siblings, and $p={100}$ phenotypic traits per individual. We study two eigenvalue configurations for $\Sigma_1$:
\begin{enumerate}[leftmargin=5pc, labelsep=0.5em]
\item[{Scenario 1.}] Linear decay: the $i$-th eigenvalue is given by $\tau_i = 1-(i-1)/p$.
\item[Scenario 2.] Step decay: $\tau_i = 1$ for $i=1,\ldots,125$, and $\tau_i = 0$ for $i=126,\ldots,250$.
\end{enumerate}
We set $\Sigma_2=\tau_e \Id_p$ and consider two noise levels in each scenario: $\tau_e=1$ and $\tau_e=5$. 

For each scenario,we compute the deterministic equivalent oracle estimator $\check d_n$, the bona fide estimator $\hat d_n$, and the isotonic bona fide estimator $\hat d_n^{\text{iso}}$ for the eigenvalues of $\Sigma_1$. We report $\ell_2$ distances to the population eigenvalues $\tau$, and $\ell_2$ distances from bona fide estimators to the oracle, with and without isotonic projection. We also compare against the MANOVA estimator $\hat{\Sigma}_{1,M}=S_n$ and the REML estimator $\hat{\Sigma}_{1,R}$, computed analytically via the Amemiya representation for balanced full-sibling designs \cite{Amemiya01051985}:
\begin{align*}
    \hat\Sigma_{1,R}= \frac{p}{n(J-1)}S_{n2}^{1/2}H\left(\frac{n-I}{I-1}S_{n2}^{-1/2}S_{n1}S_{n2}^{-1/2}\right)S_{n2}^{1/2},
\end{align*}
where $H(S)$ maps a nonnegative definite matrix $S$ with spectral decomposition $\sum_1^p\lambda_ju_ju_j^T$ into $\sum_1^p(\lambda_j-1)_+u_ju_j^T$. Note that $n(J-1)(I-1)=(n-J)(n-I)$, then 
REML agrees with MANOVA when all eigenvalues of $\frac{n-I}{I-1}S_{n1}S_{n2}^{-1}\geq 1$. 

\begin{figure}
    \centering
    \begin{subfigure}[t]{0.49\textwidth}
        \centering
        \includegraphics[trim={0cm 0cm 0cm 0cm},clip,width=\linewidth]{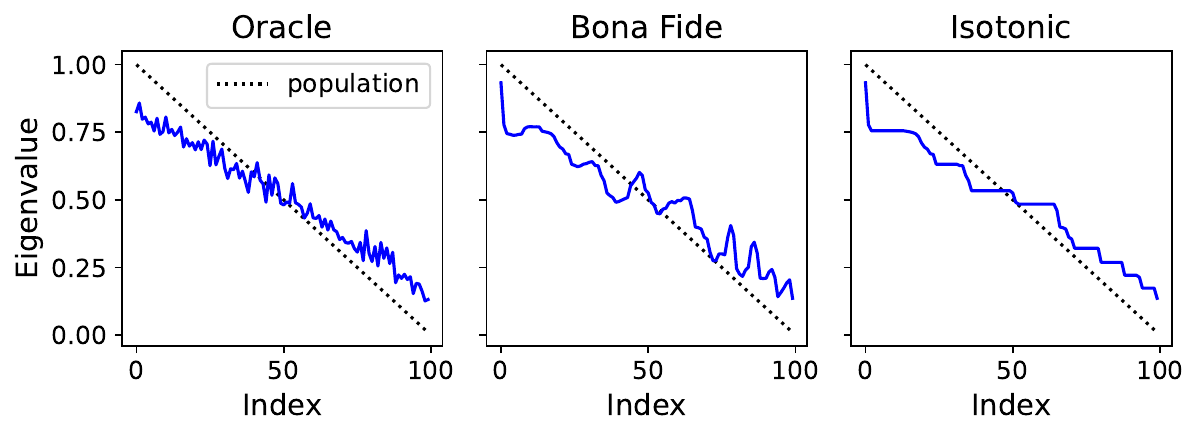}
        \caption{$\Sigma_1$ linear decay, noise level $\tau_e=1$.}
    \end{subfigure}
    \begin{subfigure}[t]{0.49\textwidth}
        \centering
        \includegraphics[trim={0cm 0cm 0cm 0cm},clip,width=\linewidth]{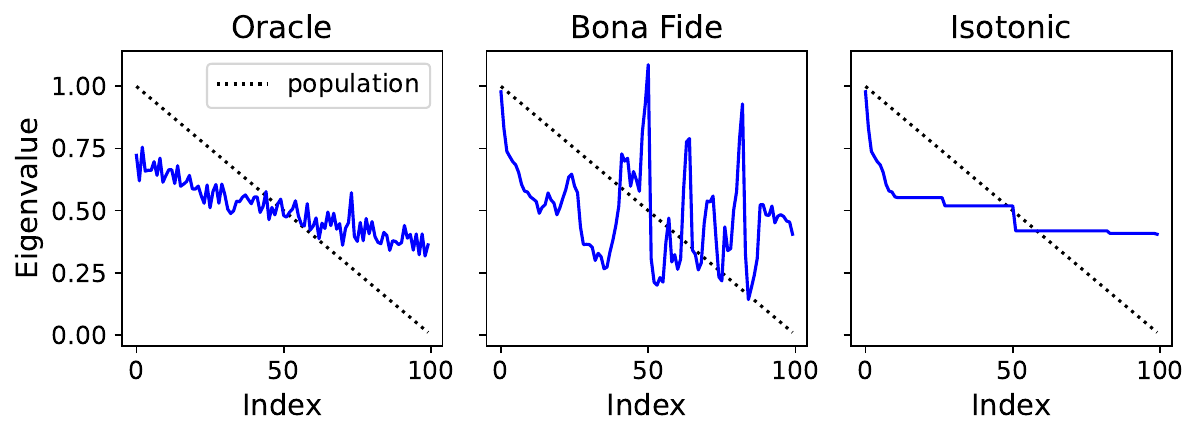}
        \caption{$\Sigma_1$ linear decay, noise level $\tau_e=5$.}
    \end{subfigure}%
    
    \begin{subfigure}[t]{0.49\textwidth}
        \centering
        \includegraphics[trim={0cm 0cm 0cm 0cm},clip,width=\linewidth]{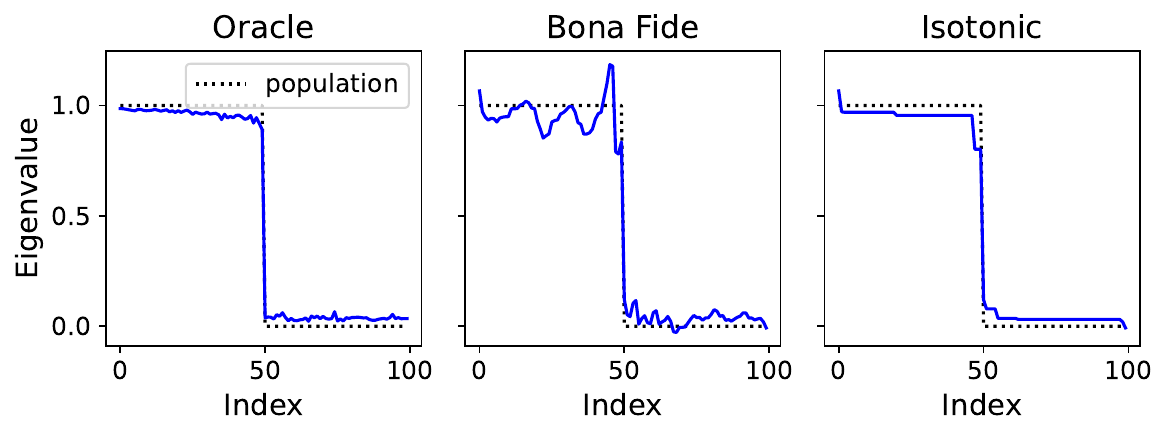}
        \caption{$\Sigma_1$ step decay, noise level $\tau_e=1$.}
    \end{subfigure}
    \begin{subfigure}[t]{0.49\textwidth}
        \centering
        \includegraphics[trim={0cm 0cm 0cm 0cm},clip,width=\linewidth]{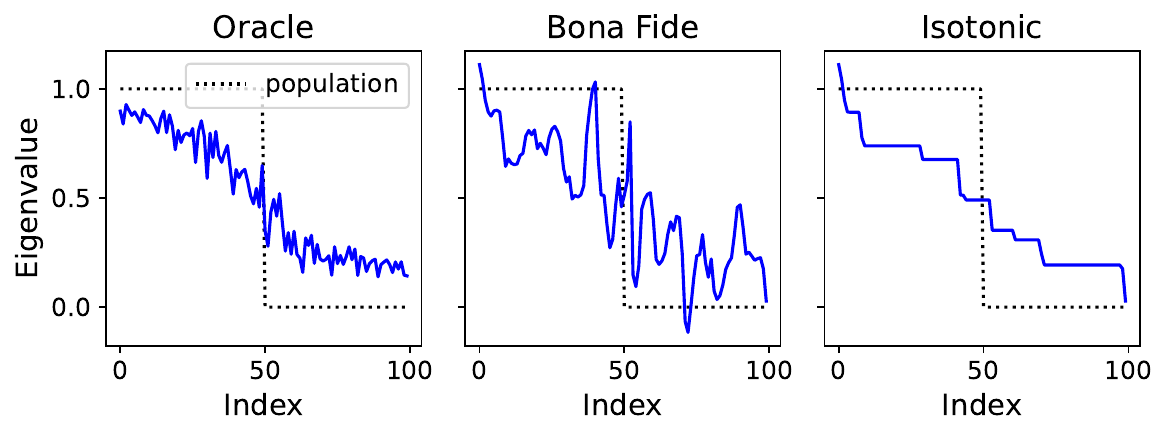}
        \caption{$\Sigma_1$ step decay, noise level $\tau_e=5$.}
    \end{subfigure}%
    \caption{Comparison of oracle, bona fide and isotonic nonparametric shrinkage estimators.}
    \label{fig:sim.full.sib}
\end{figure}

Figure \ref{fig:sim.full.sib} displays the oracle, bona fide, and isotonic bona fide nonparametric estimators under each scenario and noise level, overlaid with the population eigenvalues of $\Sigma_1$. Table \ref{tab:dist.est} reports $\ell_2$ distances to the population eigenvalues and to the oracle nonparametric estimates. Under Scenario 1 with noise level $\tau_e=1$, we also compare the proposed nonparametric shrinkage estimator with the MANOVA and REML estimators, as displayed in Figure \ref{fig:intro}.

From Figure \ref{fig:sim.full.sib}, we observe that isotonic projection substantially improves proximity both to the population eigenvalues and to the oracle shrinkage estimates $d_n$. The improvement is most pronounced at higher noise ($\tau_e=5$), where deconvolving the contribution of $\Sigma_2$ from $S_n$ is more difficult and the bona fide estimator is less stable. Consistent with Table \ref{tab:dist.est}, $\hat d_n^{\text{iso}}$ is more than three times closer to the oracle than $\hat d_n$ under linear decay and more than twice as close under step decay.

\begin{table}
    \centering
     \caption{Distance between nonparametric shrinkage estimators and the population eigenvalues of $\Sigma_1$.}
    \begin{tabular}{c c c c c c c c @{\,} c c c c c}
  & & \multicolumn{5}{c}{$\tau_e=1$} && \multicolumn{5}{c}{$\tau_e=5$} \\
  \cline{3-7} \cline{9-13}
\\[-1em]
& &  $\hat{\Sigma}_1^{\text{or}}$ & $\hat{\Sigma}_1$& $\hat{\Sigma}_1^{\text{iso}}$ &$\hat{\Sigma}_{1,M}$&$\hat{\Sigma}_{1,R}$ && $\hat{\Sigma}_1^{\text{or}}$ & $\hat{\Sigma}_1$& $\hat{\Sigma}_1^{\text{iso}}$&$\hat{\Sigma}_{1,M}$&$\hat{\Sigma}_{1,R}$\\
         \hline
        \multirow{2}{*}{Linear}&$\lVert \cdot-\tau\rVert$ & 0.992& 1.116& 1.045& 1.766& 1.735& &1.982& 2.903& 2.134& 5.973& 4.888\\
        &$\lVert \cdot-\check{d}_n\rVert$&\diagbox{}{}&0.512& 0.400& 2.645&\diagbox{}{}&&\diagbox{}{}&1.843& 0.628& 7.858	&\diagbox{}{}\\
        \multirow{2}{*}{Step}&$\lVert \cdot-\tau\rVert$ &	0.395& 0.718& 0.527& 2.855& 2.798&	&2.712& 3.321& 2.923& 6.191& 5.237\\
        &$\lVert \cdot-\check{d}_n\rVert$&\diagbox{}{}&0.590& 0.273& 2.809&\diagbox{}{}&& \diagbox{}{}1.698& 0.834& 7.183&\diagbox{}{}
    \end{tabular}
  \label{tab:dist.est}
\end{table}

\subsection{Rice data}\label{sec:rice}
We analyze gene expression data from $I=136$ accessions of the indica subspecies of \textit{Oryza sativa} assayed under drought conditions \cite{rice20}. Each accession has $J=3$ replicate measurements, giving $n=408$ observations in total. We restrict attention to the $p=100$ most heritable gene expression traits. For the $j$-th replicate of the $i$-th accession, we model the observed trait vector $y_{ij} \in \mathbb{R}^p$ as
\begin{align*}
    y_{ij} = \mu + \alpha_i + \epsilon_{ij}\in\mathbb{R}^p, \quad \alpha_i\sim N(0,{\Sigma_1}), \epsilon_{ij} \sim N(0,{\Sigma_2}),
\end{align*}
where $\mu \in \mathbb{R}^p$ is the population mean,$\alpha_i$ represents the accession-level random effect, and $\epsilon_{ij}$ is the residual error. 

We focus on estimating $\Sigma_1$, which is proportional to the additive genetic covariance matrix $G$ \cite{falconer1996introduction}. We report the isotonized nonparametric estimator and, in addition to REML and MANOVA, include method-of-moments (MoM) estimators of the eigenvalues of $\Sigma_1$ under two candidate parametric models. These MoM estimators were developed in our earlier work \cite{xie2024pargen}, and applied to the same Rice dataset (see Section 3.2 of \cite{xie2024pargen}). The quadratic model assumes that for any $x\in (0,1)$, the $\lfloor px\rfloor$-th eigenvalues of $\Sigma_1$ and $\Sigma_2$ respectively converge to
\begin{align*}
    g_r(x;\tau_r,\pi_r)=\tau_r(1-x/\pi_r)_+^2,\quad r=1,2,
\end{align*}
and the exponential model assumes
\begin{align*}
    g_r(x;\tau_{1,r},\tau_{2,r})=\tau_{1,r}\exp(-\tau_{2,r}x),\quad r=1,2.
\end{align*}

\begin{figure}
    \centering
    \begin{subfigure}[t]{\textwidth}
        \centering
        \includegraphics[trim={0cm 0cm 0cm 0cm},clip,width=1\linewidth]{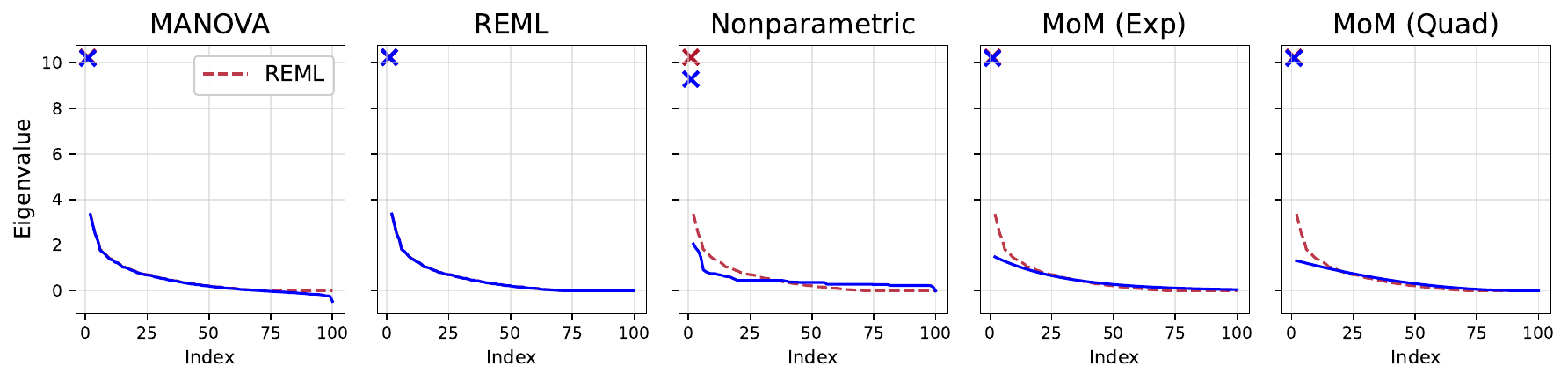}
        \caption{Comparison of REML with various estimators of the eigenvalues of $\Sigma_1$. }
    \end{subfigure}

   \begin{subfigure}[t]{\textwidth}
        \centering
        \includegraphics[trim={0cm 0cm 0cm 0cm},clip,width=1\linewidth]{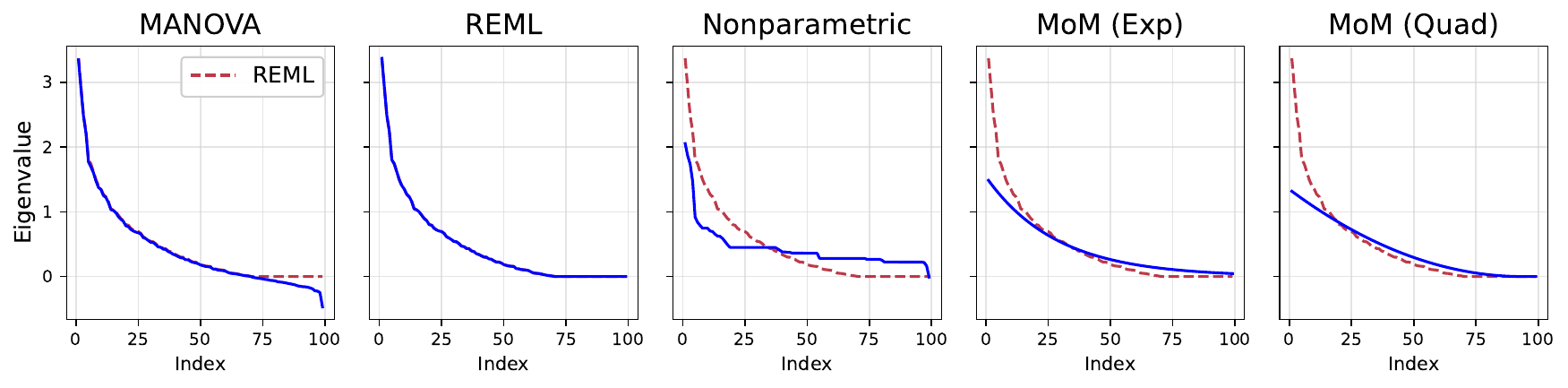}
        \caption{Same as (a), excluding the spike.}
    \end{subfigure}

    \caption{Rice data. Comparison of REML with various estimators of the eigenvalues of $\Sigma_1$.     }\label{fig:rice}
\end{figure}

Figure~\ref{fig:rice} compares estimated eigenvalues of $\Sigma_1$ across the five methods. REML and MANOVA are broadly similar, except that MANOVA produces some negative eigenvalues, whereas REML is constrained to be positive semidefinite. Relative to REML and MANOVA, the nonparametric shrinkage estimator and both MoM estimators pull the spectrum toward its center by increasing smaller eigenvalues and decreasing larger ones, thereby reducing overall dispersion. REML and the quadratic MoM fit both imply a nontrivial estimated null space, whereas the nonparametric estimator does not. This is expected: unless a null-space eigenvector $u_i$ is estimated exactly by MANOVA, the oracle target $u_i^\top \Sigma_1 u_i$ is generally nonzero. At the the lower end of the spectrum, the behavior of the nonparametric estimator resembles that of MANOVA. This is consistent with the fact that it is constructed by preserving the eigenvectors while shrinking the eigenvalues of MANOVA. Overall, these patterns suggest that REML may overstate upper-tail and lower-tail eigenvalues, potentially inflating inferred high evolvabilities and evolutionary constraints.

\section*{Acknowledgements}
The work of both authors was supported in part by NIH grant R01 GM134483.

\newpage
\printbibliography
\newpage
\appendix

\section{Proof of Lemma \ref{lem:inversion.borel}}\label{sec:ar.br.borel}
We prove the two parts of Lemma \ref{lem:inversion.borel} separately. Part (1), concerning $\tilde{a}_n^{(r)}$, follows by adapting Lemma 2.2 and Lemma A.3 of \cite{xie2024clt}. The remainder of this section is devoted to proving part (2), concerning $\tilde{b}_n^{(r)}$. The proof proceeds in two steps. We first establish that $\tilde{b}_n^{(r)}(z)$ is the Stieltjes transform of a positive Borel measure $\tilde{\mu}_{n,b}^{(r)}$ on $\mathbb{R}$, and then verify that $\tilde{\mu}_{n,b}^{(r)}$ serves as a deterministic equivalent for $\mu_{n,b}^{(r)}$ defined in \eqref{eq:mu.n.b}.

\subsection{$\tilde{b}_n^{(r)}$ as Stieltjes transform}
Our proof relies on the following characterization of Stieltjes transforms.
\begin{lem}\label{lem:stieltjes.characterization}
    Suppose $G:\mathbb{C}^+\rightarrow\mathbb{C}^+$ is analytic and satisfies $\lim_{y\rightarrow\infty}iyG(iy)=-c$ for some $0<c<\infty$. Then there exists a unique positive Borel measure $\nu$ on $\mathbb{R}$ with $\nu(\mathbb{R})=c$ such that
    \begin{align*}
        G(z) = \int_{\mathbb{R}}\frac{1}{t-z}d\nu(t).
    \end{align*}
\end{lem}
\begin{proof}
    This follows from Lemma 2 in \cite{Geronimo2003}.
\end{proof}

It therefore suffices to verify that $\tilde{b}_n^{(r)}(z)$ satisfies the conditions of Lemma~\ref{lem:stieltjes.characterization}. We first establish that $\Im [{b}_n^{(r)}(z)]>0$ for any $z\in\mathbb{C}^+$, which implies $\Im [\tilde{b}_n^{(r)}(z)]>0$ by definition given in \eqref{eq:tilde.b.n.r}. Recall from \eqref{eq:b.n.r} that
\begin{align*}
    p{b}_n^{(r)}&=-\Tr ((L^{-1}+\sum_{s=1}^k\tilde{a}_n^{(s)}\tilde{R}_s^T\tilde{R}_s)^{-1}\tilde{R}_r^T\tilde{R}_r)
\end{align*}
Define $S=(L^{-1}+\sum_{s=1}^k\tilde{a}_n^{(s)}\tilde{R}_s^T\tilde{R}_s)^{-1}$. Then 
\begin{align*}
     pb_n^{(r)}&=-\Tr (S\tilde{R}_r^T\tilde{R}_rS^*(L^{-1}+\sum_{s=1}^k\overline{\tilde{a}_n^{(s)}}\tilde{R}_s^T\tilde{R}_s))\\
     &=-\Tr (S\tilde{R}_r^T\tilde{R}_rS^*L^{-1})-{\textstyle\sum}_{s=1}^k\overline{\tilde{a}_n^{(s)}}\Tr (S\tilde{R}_r^T\tilde{R}_rS^*\tilde{R}_s^T\tilde{R}_s),
\end{align*}
and consequently
\begin{align*}
   p\Im b_n^{(r)} = {\textstyle\sum}_{s=1}^k\Im{\tilde{a}_n^{(s)}}\Tr (S\tilde{R}_r^T\tilde{R}_rS^*\tilde{R}_s^T\tilde{R}_s).
\end{align*}
We prove $\Im \, b_n^{(r)}>0$ by contradiction. From Lemma \ref{lem:deteqv} part (1), we have that $\Im \,{\tilde{a}_n^{(s)}}>0$ for each $s$. Since $\Tr (S\tilde{R}_r^T\tilde{R}_rS^*\tilde{R}_s^T\tilde{R}_s)\geq 0$, the assumption $\Im \, b_n^{(r)}=0$ implies $\tilde{R}_sS\tilde{R}_r^T=0$, and as a result $R_sSR_r^T=0$ for all $s$. In Supplementary Appendix \ref{app:deteqv}, equation \eqref{eq:br.sherman.morrison.2}, we show that for $F$ defined in \eqref{eq:f} and there exists an invertible matrix $M$ such that,
\begin{align*}
    M^{-1}F = RSR^T.
\end{align*}
If $R_sSR_r^T=0$ for all $s$, then $RSR_r^T=0$, so the $I_r$ columns of $M^{-1}F$ corresponding to index $r$ vanish. Denoting these columns of $F$ by $F_{\cdot r}$, we obtain
\begin{align*}
    M^{-1}F_{.r}=0\Rightarrow F_{.r}=0\Rightarrow F_{rr}=\frac{I_r}{p}U_r^TBU_r=0,
\end{align*}
contradicting the assumption $U_r^TBU_r\neq 0$. We conclude that 
$$
\Im \, b_n^{(r)}(z)=\Im \,\tilde{b}_n^{(r)}(z)>0.
$$

Next, we show that $i\tilde{b}_n^{(r)}(iy)$ converges as $y\rightarrow\infty$. By definitions of $b_n^{(r)}$ 
and $\tilde{b}_n^{(r)}(z)$ in \eqref{eq:b.n.r} and \eqref{eq:tilde.b.n.r}, we have

    \begin{align*}
    \tilde{b}_n^{(r)}(z)
    &=-\frac{1}{p}\Tr ((\Id _{l_0}+\sum_{s=1}^k\tilde{a}_n^{(s)}\tilde{R}_s^T\tilde{R}_sL)^{-1}\tilde{R}_r^T\tilde{R}_rL)+\frac{1}{p}\Tr (\tilde{R}_r^T\tilde{R}_rL)\\
    &=\frac{1}{p}\Tr \left(\left[\Id _{l_0}-(\Id _{l_0}+\sum_{s=1}^k\tilde{a}_n^{(s)}\tilde{R}_s^T\tilde{R}_s)^{-1}\right]\tilde{R}_r^T\tilde{R}_rL\right),\\
    &=\frac{1}{p}\Tr \left(\sum_{t=1}^k\tilde{a}_n^{(t)}\tilde{R}_t^T\tilde{R}_tL(\Id _{l_0}+\sum_{s=1}^k\tilde{a}_n^{(s)}\tilde{R}_s^T\tilde{R}_sL)^{-1}\tilde{R}_r^T\tilde{R}_rL\right)\\
    &=\sum_{t=1}^k\frac{1}{p}\Tr \left(\tilde{a}_n^{(t)}\tilde{R}_t^T\tilde{R}_tL(\Id _{l_0}+\sum_{s=1}^k\tilde{a}_n^{(s)}\tilde{R}_s^T\tilde{R}_sL)^{-1}\tilde{R}_r^T\tilde{R}_rL\right).
\end{align*}
As a result, using the identity $iy\tilde{a}_n^{(r)}(iy)\rightarrow-p^{-1}\Tr (\Sigma_r)$ as $y\rightarrow\infty$ given in part 1, we have
\begin{align*}
    iy\tilde{b}_n^{(r)}(iy)&=\sum_{t=1}^k\frac{1}{p}\Tr \left(i\tilde{a}_n^{(t)}(iy)\tilde{R}_t^T\tilde{R}_tL(\Id _{l_0}+\sum_{s=1}^k\tilde{a}_n^{(s)}(iy)\tilde{R}_s^T\tilde{R}_sL)^{-1}\tilde{R}_r^T\tilde{R}_rL\right)\\
    &\rightarrow-\frac{1}{p^2}\sum_{t=1}^k\Tr (\Sigma_r)\Tr \left(\tilde{R}_r^T\tilde{R}_rL\tilde{R}_t^T\tilde{R}_tL\right).
\end{align*}
\subsection{$\tilde\mu_{n,b}^{(r)}$ as deterministic equivalent}
    
Our goal is to show that
\begin{align*}
    \tilde{b}_n^{(r)}(z)-m_{n,b}^{(r)}(z)\rightarrow_{a.s.}0,\quad m_{n,b}^{(r)}=\frac{1}{p}\Tr ((S_n-z\Id )^{-1}S_n^{(r)}).
\end{align*}
Define $\alpha=[I_1^{-1/2}\alpha_1^T|\ldots|I_k^{-1/2}\alpha_k^T]^T$, along with $U$, $F$, $R$ defined in \eqref{eq:f}, then we can alternatively write $S_n$ in (\ref{eq:sos}) as 
\begin{align*}
    S_n &=\frac{1}{p}\alpha^TU^T BU\alpha=\alpha^TRLR^T\alpha^T.
\end{align*}
Define
\begin{align*}
    \mathcal{S}_n&=R^T\alpha\alpha^TRL\in\mathbb{R}^{l_0\times l_0}.
\end{align*}
In Supplementary Appendix \ref{app:bn.bn.tilde}, we show that
\begin{align}\label{eq:bn.bn.tilde}
    \tilde{b}_n^{(r)}(z)-m_{n,b}^{(r)}(z) = b_n^{(r)}-(-p^{-1})z\Tr ((\mathcal{S}_n-z\Id )^{-1}\tilde{R}_r^T\tilde{R}_rL).
\end{align}
As a result, in the following we show that the latter vanishes almost surely.

Let $T_r=\tilde{R}_rL$, then
\begin{align*}
    \mathcal{S}_n=_d\sum_{r=1}^k\sum_{s=1}^k\tilde{R}_r^T(p^{-1/2}X_r)\Sigma_r^{1/2}\Sigma_s^{1/2}(p^{-1/2}X_s^T)T_s.
\end{align*}

Below, we generalize the proof strategy in Section 4 of \cite{fan2017eigenvalue} using free probability theory to proof the following lemma. 
\begin{lem}\label{lem:mathcal.Sn}
    Under Assumption \ref{assum:minimum}, for each $z\in\mathbb{C}^+$, and $r=1,\ldots,k$, we have
\begin{align*}
    b_n^{(r)}-(-p^{-1})z\Tr ((\mathcal{S}_n-z\Id )^{-1}\tilde{R}_r^T\tilde{R}_rL)\rightarrow 0.
\end{align*}
\end{lem}

\subsubsection{Proof of Lemma \ref{lem:mathcal.Sn}}
We refer to Section 3 in \cite{fan2017eigenvalue} for background on operator-valued free probability theory. In particular, let a noncommutative probability space $(\mathcal{A},\tau)$, a $\mathcal{B}$-valued probability space $(\mathcal{A},\mathcal{B},\mathbf{F}^{\mathcal{B}})$ with conditional expectation $\mathbf{F}^{\mathcal{B}}$, the $\mathcal{B}$-\textit{valued} $\mathcal{R}$-\textit{transform} of $a\in\mathcal{A}$, for $b\in\mathcal{B}$ with $16\lVert b\rVert\leq 1/\lVert a\rVert$, denoted $\mathcal{R}_a^{\mathcal{B}}(b)$, and the $\mathcal{B}$-\textit{valued} \textit{Cauchy-transform} of $a\in\mathcal{A}$, for invertible $b\in\mathcal{B}$ with $\lVert b^{-1}\rVert\leq 1/\lVert a\rVert$, denoted $G_a^{\mathcal{B}}(b)$, be defined as in \cite{fan2017eigenvalue}. By (3.15) in \cite{fan2017eigenvalue}, the Cauchy-transform satisfies the upper bound,
\begin{align}\label{eq:cauchy.upper}
    G_a^{\mathcal{B}}(b)\leq \frac{\lVert b^{-1}\rVert}{1-\lVert a \rVert \lVert b^{-1}\rVert}.
\end{align}

We generalize the proof strategy in Section 4 of \cite{fan2017eigenvalue} to our case. Many steps remain analogous to those in the original work, we include them here for completeness. We extend the notations of \cite{fan2017eigenvalue}. Fix $k\geq 1$, and let $l_0,n_1,\ldots,n_k,m_1,\ldots,m_k\in\mathbb{N}$, and denote $n_+=\sum_{r=1}^kn_r$. Let ${F}\in\mathbb{C}^{n_+\times n_+}$ be deterministic with $F^*=F$, and denote by $F_{rs}$ its $(r,s)$ submatrix. For $r=1,\ldots,k$, let $H_r,T_r\in\mathbb{C}^{m_r\times l_0}$ be deterministic, and let $G_r$ be independent random matrices such that either ${G}_r\in \mathbb{R}^{n_r\times m_r}$ with $(G_r)_{ij}\iidsim N(0,n_r^{-1})$ or ${G}_r\in \mathbb{C}^{n_r\times m_r}$ with $\Im(G_r)_{ij},\Re(G_r)_{ij}\iidsim N(0,(2n_r)^{-1})$. Define 
\begin{align*}
    W=\sum_{r,s=1}^kH_r^*{G}_r^*F_{rs}{G}_s{T}_s\in \mathbb{C}^{l_0\times l_0},
\end{align*}
with empirical spectral measure $\mu_W$.  
{Taking $n_r=p$, $m_r=I_r$, $n_+=\sum_rn_r=kp$, $F_{rs}=\Sigma_r^{1/2}\Sigma_s^{1/2}$, $H_r=\tilde{R}_r$, $T_r=\tilde{R}_rL$, and $G_r$ to be real, then we have $W=_d\mathcal{S}_n$.}

Following the approach in \cite{fan2017eigenvalue}, we prove Lemma \ref{lem:mathcal.Sn} by first defining a free deterministic equivalent $\omega$ for $W$, and then show that the Stieltjes transform of $\omega$ satisfies the system of equations given in the lemma.

Before we proceed with the proof, we introduce some notations and definitions that will be used to define the free deterministic equivalent $\omega$.
Consider the transformations $T_r\rightarrow O_r^TT_rO_0$, $H_r\rightarrow O_r^TH_rO_0$, $F_{rs}\rightarrow O_{k+r}^TF_{rs}O_{k+s}$ for independent Haar-distributed orthogonal matrices $O_0,\ldots,O_{2k}$ of the appropriate sizes. It is easy to check that $\mu_W$ remains invariant in law under these transformations. Hence it suffices to prove Lemma \ref{lem:free.prob} with $T_r$, $H_r$, and $F_{rs}$ replaced by the randomly rotated matrices, which we continue to denote by $T_r$, $H_r$, and $F_{rs}$.
 
Let $N=l_0+\sum_rm_r+\sum_rn_r$, and embed the matrices as
simple elements of $\mathbb{C}^{N\times N}$ in the following regions of the block-matrix decomposition corresponding to $\mathbb{C}^N=\mathbb{C}^{l_0}\oplus\mathbb{C}^{m_1}\oplus\ldots\oplus\mathbb{C}^{m_k}\oplus\mathbb{C}^{n_1}\oplus\ldots\oplus\mathbb{C}^{n_k}$:
\begin{align*}
    \begin{bmatrix}
        W&H_1^*&\ldots&H_k^*&&&\\
        T_1&&&&G_1^*&&\\
        \vdots&&&&&\ddots&\\
        T_k&&&&&&G_k^*\\
        &G_1&&&F_{11}&\ldots&F_{1k}\\
        &&\ddots&&\vdots&\ddots&\vdots\\
        &&&G_k&F_{k1}&\ldots&F_{kk}
    \end{bmatrix}
\end{align*}
Denote by $P_0,\ldots,P_{2k}$ diagonal projections corresponding to the above decomposition, and by $\tilde{W}$, $\tilde{G}_r$, $\tilde{H}_r$, $\tilde{T}_r$, $\tilde{F}_{rs}\in\mathbb{C}^{N\times N}$ the embedded matrices. Then $\tilde{W}$, $\tilde{G}_r$, $\tilde{H}_r$, $\tilde{T}_r$, $\tilde{F}_{rs}$ are simple elements of
the rectangular space $(\mathbb{C}^{N\times N},N^{-1}\Tr ,P_0,\ldots,P_{2k})$, and the $k+2$ families, $\{\tilde{F}_{rs}\}$, $\{\tilde{H}_r,\tilde{T}_r\}$, $\tilde{G}_1,\ldots,\tilde{G}_k$ are independent of each other and are block-orthogonally invariant.

For the approximating free model, consider a second (N-dependent) rectangular
space $(\mathcal{A},\tau,p_0,\ldots,p_{2k})$ with deterministic elements $f_{rs},g_r,h_r,t_r\in\mathcal{A}$, such that the following hold:
\begin{enumerate}
    \item $p_0,\ldots,p_{2k}$ have traces
    \begin{align*}
        \tau(p_0)=l_0/N, \quad \tau(p_r)=m_r/N, \quad \tau(p_{k+r})=n_r/N, \quad \text{for all } r=1,\ldots,k.
    \end{align*}
    \item $f_{rs},g_r,h_r,t_r$ are simple elements such that for all $r,s\in\{1,\ldots,k\}$,
    \begin{align*}
        p_{k+r}f_{rs}p_{k+s}=f_{rs}, \quad p_{k+r}g_rp_r=g_r, \quad p_rh_rp_0=h_r, \quad p_rt_rp_r=t_r.
    \end{align*}
    \item $\{f_{rs}:1\leq r,s\leq k\}$ has the same joint $\mathcal{D}$-law as $\{\tilde{F}_{rs}:1\leq r,s\leq k\}$, $\{h_r,t_r:1\leq r\leq k\}$ has the same joint $\mathcal{D}$-law as $\{\tilde{H}_r,\tilde{T}_r:1\leq r\leq k\}$. That is, for any $r\in\{0,\ldots,2k\}$, and any noncommutative $\ast$-polynomials $\mathcal{Q}_1,\mathcal{Q}_2$ with coefficients in $\langle P_0,\ldots,P_{2k}\rangle$,letting $q_1,q_2$ denote the corresponding $\ast$-polynomials with coefficients in $\langle p_0,\ldots,p_{2k}\rangle$,
    \begin{align}
        \tau_r[q_1(f_{uv}:u,v\in\{1,\ldots,k\})]&=N_r^{-1}\Tr _r[\mathcal{Q}_1(\tilde{F}_{uv}:u,v\in\{1,\ldots,k\})],\\
        \tau_r[q_2(h_{u},t_{v}:u,v\in\{1,\ldots,k\})]&=N_r^{-1}\Tr _r[\mathcal{Q}_2(\tilde{H}_u,\tilde{T}_v:u,v\in\{1,\ldots,k\})],\label{eq:d.law.H}
    \end{align}
    where $N_0=l_0$, $N_r=m_r$ for $r=1,\ldots,k$ and $N_{r}=n_r$ for $r=k+1,\ldots,2k$.
    \item For each $r$, $g^*g_r$ has Marcenko–Pastur law with parameter with parameter $\lambda=m_r/n_r$. That is, for 
    \begin{align*}
        \nu_{\lambda}(x)=\frac{1}{2\pi}\frac{\sqrt{(\lambda_+- x)(x-\lambda_-)}}{\lambda x}\mathbbm{1}_{[\lambda_-,\lambda_+]}(x),\quad \lambda_{\pm}=(1\pm\sqrt{\lambda})^2,
    \end{align*}
    we have
    \begin{align}\label{eq:g.r.moment}
        \tau_r((g_r^*g_r)^l)=\int x^l\nu_{m_r/n_r}(x)dx\quad \text{for all } l\geq 0.
    \end{align}
    \item The $k+2$ families $\{f_{rs}\}$, $\{h_r,t_r\}$, $g_1,\ldots,g_k$ are free with amalgamation over $\langle p_0,\ldots,p_{2k}\rangle$.
\end{enumerate}
The following lemma establishes existence of this model as a von Neumann
algebra. We refer to \cite[Appendix D]{fan2017eigenvalue} for the definition of von Neumann algebras with norm $\lVert \cdot \rVert$, and norm-continuous trace $\tau$ satisfying $|\tau(a)|\leq \lVert a\rVert$.
\begin{lem}\label{lem:deteqv.free.prob}
    Under the conditions of Lemma \ref{lem:free.prob}, there exists a ($N$-dependent) rectangular probability space $(\mathcal{A},\tau,p_0,\ldots,p_{2k})$ such that:
    \begin{enumerate}
        \item $\mathcal{A}$ is a von Neumann algebra and $\tau$ is a positive, faithful, normal trace.
        \item $\mathcal{A}$ contains elemts $f_{rs},g_r,h_r,t_r$ for $r,s\in\{1,\ldots,k\}$ that satisfy the above conditions. Furthermore, the von Neumann subalgebras $\langle \mathcal{D},\{f_{rs}\}\rangle_{W^*}$, $\langle \mathcal{D},\{h_r,t_r\}\rangle_{W^*}$, $\langle \mathcal{D},\{g_1\}\rangle_{W^*}$, $\langle \mathcal{D},\{g_k\}\rangle_{W^*}$ are free over $\mathcal{D}$.
       \item There exist a constant $C>0$ such that $\lVert f_{rs}\rVert,\lVert h_r\rVert,\lVert t_r\rVert,\lVert g_r\rVert\leq C$ for all $N$ and all $r,s$. 
    \end{enumerate}
\end{lem}
\begin{proof}
    This is a straightforward extension of the proof of Lemma 4.2 in \cite{fan2017eigenvalue}.
\end{proof}
Following the convention in \cite{fan2017eigenvalue}, we also identify $\mathbb{C}\subset\mathcal{A}$ as a subalgebra via the inclusion map $z\rightarrow z{1}_{\mathcal{A}}$, and we write $1$ for $1_{\mathcal{A}}$ and $z$ for $z{1}_{\mathcal{A}}$.

{Since $\mathcal{S}_n$ is no longer Hermitian, with $H_r\neq T_r$,
Lemma~4.3 in \cite{fan2017eigenvalue} does not apply directly. We
establish the following extension.}
\begin{lem}[Generalization of Lemma 4.3 in \cite{fan2017eigenvalue}]\label{lem:general.lem.4.3}
Let $(\mathcal{A},\tau,q_0,q_1,\ldots,q_k)$ be a rectangular probability space,
where $\mathcal{A}$ is von Neumann and $\tau$ is positive, faithful and normal. Let $\mathcal{D}=\langle q_0,q_1,\ldots,q_k),\rangle$, let $\mathcal{B},\mathcal{C}\subset\mathcal{A}$ be von Neumann subalgebras containing $\mathcal{D}$ that are free over $\mathcal{D}$ and let $F^{\mathcal{D}}:\mathcal{A}\rightarrow \mathcal{D}$ and $F^{\mathcal{C}}:\mathcal{A}\rightarrow \mathcal{C}$ be the $\tau$-invariant conditional
expectations.

Let $b_{rs}\in\mathcal{B}$ and 
{$c_r,d_r\in\mathcal{C}$} for $1\leq r,s\leq k$ be such that $q_rb_{rs}q_s=b_{rs}$, 
{$q_rc_r=c_r$, $q_rd_r=d_r$}, $\lVert b_{rs}\rVert\leq C$, and 
{$\lVert c_r\rVert,\lVert d_r\rVert\leq C$} for some constant $C>0$. Define 
{$a=\sum_{r,s=1}^kc_r^*b_{rs}d_s$} and $b=\sum_{r,s=1}^kb_{rs}$. Then, for $e\in\mathcal{C}$ with $\lVert e\rVert$ sufficiently small,
\begin{align*}
    \mathcal{R}_a^{\mathcal{C}}(e)=\sum_{r=1}^kc_r^*d_s\tau_r(\mathcal{R}_b^{\mathcal{D}}(\sum_{s=1}^k\tau_s(d_sec_s^*)q_s)).
\end{align*}
\end{lem}
\begin{proof}
    It is easy to verify that the proof of Lemma 4.3 in \cite{fan2017eigenvalue} still holds under this setting, with $c_s$ in Lemma 4.3 replaced by $d_s$ in our case.
\end{proof}

For $x=(x_1,\ldots,x_k)$, let $D(x) = \text{diag}(x_1\Id _{n_1},\ldots,x_k\Id _{n_k})\in\mathbb{C}^{n_+\times n_+}$. For any $F\in\mathbb{C}^{n_+\times n_+}$, let $\Tr_r F$ denote the trace of its $(r,r)$ block in the $k\times k$ block decomposition corresponding to $\mathbb{C}^{n^+}=\mathbb{C}^{n_1}\oplus\ldots \oplus\mathbb{C}^{n_k}$.

\begin{lem}\label{lem:free.prob}
    Under the conditions of Lemma \ref{lem:mathcal.Sn}, let $(\mathcal{A},\tau,p_0,\ldots,p_{2k})$ and $f_{rs},g_r,h_r,t_r$ be as in Lemma \ref{lem:deteqv.free.prob}, and let $\omega=\sum_{r,s=1}^kh_r^*g_r^*f_{rs}g_st_s$. Then for a constant $C_0>0$, defining $\mathbb{D}:=\{z\in\mathbb{C}^+:|z|>C_0\}$, there exists z-dependent values $x_1,\ldots,x_k\in{\mathbb{C}}$ and $y_1,\ldots,y_k\in\mathbb{C}$ that satisfy, for $r=1,\ldots,k$, the equations
        \begin{align*}
            x_r
            &=-\frac{1}{n_r}\Tr \left(\left(\Id _{l_0}+{\textstyle\sum}_{s=1}^ky_rH_s^*T_s\right)^{-1}H_r^*T_r\right),\\
            y_r& = -\frac{1}{n_r}\Tr _r\left((z\Id _{n_+}+FD(x))^{-1}F\right).
        \end{align*}
    Moreover, for each $r=1,\ldots,k$, the function $x_r$ satisfies 
        \begin{align*}
            x_r-(-n_r^{-1})z\Tr ((W-z\Id )^{-1}H_r^*T_r)\rightarrow 0.
        \end{align*}
\end{lem}

\begin{proof}[Proof of Lemma \ref{lem:mathcal.Sn}]
    {Taking $n_r=p$, $m_r=I_r$, $n_+=\sum_rn_r=kp$, $F_{rs}=\Sigma_r^{1/2}\Sigma_s^{1/2}$, $H_r=\tilde{R}_r$, $T_r=\tilde{R}_rL$, and $G_r$ to be real, then we have $W=_d\mathcal{S}_n$.}
     Applying the matrix identity $(AB +\Id)^{-1}A = A(BA + \Id)^{-1}$ as in the proof of Lemma \ref{lem:deteqv} in Appendix \ref{app:deteqv}, we can equivalently write the above system of equations as
    \begin{align*}
        x_r&=-\frac{1}{p}\Tr \left((\Id _{l_0}+\sum_{s=1}^ky_s\tilde{R}_s^T\tilde{R}_sL)^{-1}\tilde{R}_r^T\tilde{R}_rL\right)\\
        y_r
        &= -\frac{1}{p}\Tr ((z\Id _p+{\textstyle\sum}_{s=1}^kx_s\Sigma_s)^{-1}\Sigma_r).
    \end{align*}
    By Lemma \ref{lem:deteqv}, we know that for each $z\in\mathbb{C}^+$, there exists unique z-dependent values $x_1,\ldots,x_k\in\overline{\mathbb{C}^+}$, $y_1,\ldots,y_k\in\mathbb{C}^+\cup\{0\}$ that satisfy the above system, with $x_r=b_n^{(r)}$, $y_r=\tilde{a}_n^{(r)}$. Then for each $z\in\mathbb{D}$, we have
    \begin{align*}
        b_n^{(r)}-(-p^{-1})z\Tr ((\mathcal{S}_n-z\Id )^{-1}\tilde{R}_r^T\tilde{R}_rL)\rightarrow 0.
    \end{align*}
    By \eqref{eq:tilde.b.n.r}, and that $\tilde{b}_n^{(r)}$ defines a Stieltjes transform, we have that $ b_n^{(r)}-(-I_r^{-1})z\Tr ((\mathcal{S}_n-z\Id )^{-1}R_r^TR_rL)$ is bounded over $\{z\in\mathbb{C}^+:\Im z>\epsilon\}$ for any $\epsilon>0$, then Lemma C.1 in \cite{fan2017eigenvalue} implies that 
\begin{align*}
    b_n^{(r)}-(-I_r^{-1})z\Tr ((\mathcal{S}_n-z\Id )^{-1}R_r^TR_rL)\rightarrow_{a.s.}0.
\end{align*}
for all $z\in\mathbb{C}^+$. 
\end{proof}

\subsubsection{Proof of Lemma \ref{lem:free.prob}}

Define the von Neumann subalgebras $\mathcal{D}=\langle p_r:0\leq r\leq 2k\rangle$, $\mathcal{F}=\langle \mathcal{D},\{f_{rs}\}\rangle_{W^*}$, $\mathcal{G}=\langle \mathcal{D},\{g_r\}\rangle_{W^*}$, $\mathcal{H}=\langle \mathcal{D},\{h_r,t_r\}\rangle_{W^*}$. Denote by $F^{\mathcal{D}}$ and $\mathcal{R}^{\mathcal{D}}$, and $G^{\mathcal{D}}$ the $\tau$-invariant conditional expectation onto $\mathcal{D}$ and the $\mathcal{D}$-valued $\mathcal{R}$-transform and Cauchy transform, and similarly for $\mathcal{F}$, $\mathcal{G}$, $\mathcal{H}$.

 We first relate the $\mathcal{D}$-valued $\mathcal{R}$-transform of $\omega=\sum_{r,s=1}^kh_r^*g_r^*f_{rs}g_st_s$ to that of $\nu:=\sum_{r,s=1}^kg_r^*f_{rs}g_s$. We apply Lemma \ref{lem:general.lem.4.3} with $q_0=p_0+\sum_{r=k+1}^{2k}p_r$, $q_r=p_r$ for $r=1,\ldots,k$, $\mathcal{C}=\mathcal{H}$ and $\mathcal{B}=\langle\mathcal{F},\mathcal{G}\rangle$. Then for $c\in\mathcal{H}$ sufficiently small,
\begin{align}\label{eq:R.omega.H}
    \mathcal{R}_{\omega}^{\mathcal{H}}(c)=\sum_{r=1}h_r^*t_r\tau_r(\mathcal{R}_{\nu}^{\mathcal{D}}(\sum_{s=1}^kp_s\tau_s(t_sch_s^*))).
\end{align}
By Lemma \ref{lem:deteqv.free.prob}, we have that $\lVert\omega\rVert\leq C$. Then for $C_0$ large enough and invertible $d\in\mathcal{D}$ with $\lVert d^{-1}\rVert$ sufficiently small,  $G_{\omega}^{\mathcal{H}}(d)$ is well defined and arbitrarily small by \eqref{eq:cauchy.upper}. For each $r=1,\ldots,k$, further define 
\begin{align}
    \chi_r(d)&:=\tau_r(t_r{G}_{\omega}^{\mathcal{H}}(d)h_r^*),\label{eq:chi.r}\\
    \upsilon_r(d)&:=\tau_r(\mathcal{R}_{\nu}^{\mathcal{D}}(\textstyle{\sum}_{s=1}^kp_s\chi_s(d))).
\end{align}
Then by Proposition 3.12 in \cite{fan2017eigenvalue} and \eqref{eq:R.omega.H} with $c=G_{\omega}^{\mathcal{H}}(d)$, we have
\begin{align*}
    G_{\omega}^{\mathcal{H}}(d) &= (d-\mathcal{R}_{\omega}^{\mathcal{H}}(G_{\omega}^{\mathcal{H}}(d)))^{-1}= (d-\textstyle{\sum}_{r=1}^kh_r^*t_r\upsilon_r(d))^{-1}.
\end{align*}
Plugging this in the definition of $\chi_r$ in \eqref{eq:chi.r}, we have
\begin{align*}
    \chi_r(d)&=\tau_r\left(t_r(d-{\textstyle\sum}_{s=1}h_s^*t_s\upsilon_s(d))^{-1}h_r^*\right).
\end{align*}
Denote $d=d_0p_0+\ldots+d_{2k}p_{2k}$ for $d_0,\ldots,d_{2k}\in\mathcal{C}$. As $d$ is invertible, we have $d^{-1}=d_0^{-1}p_0+\ldots+d_{2k}^{-1}p_{2k}$.
For any $x\in\mathcal{A}$ that commutes with $\mathcal{D}$, 
\begin{align*}
    (d-x)^{-1}=\sum_{l\geq 0}d^{-1}(xd^{-1})^l=\sum_{l\geq 0}x^ld^{-l-1}.
\end{align*}
{
Noting that $p_r=p_r^2$ and that $\mathcal{D}$ commutes with itself, then for any $a,b\in\mathcal{A}$ satisfying $p_rap_0=a$, $p_rbp_0=b$, we have
\begin{align*}
    \tau_r(a(d-x)^{-1}b^*)=\frac{N}{m_r}\sum_{l\geq 0}\tau((p_0dp_0)^{-l-1}p_0x^lb^*p_rp_rap_0)=\frac{m_0}{m_r}\sum_{l\geq 0}\frac{\tau_0(x^lb^*a)}{d_0^{l+1}}.
    \end{align*}
{Note that $h_r^*t_r$ commutes with $\mathcal{D}$ and $p_rh_rp_0=h_r$, $p_rt_rp_0=t_r$,} then let $x = {\textstyle\sum}_{r=1}h_r^*t_r\upsilon_r(d)$, $a=t_r$, $b=h_r$, for $z\in\mathbb{C}^+$, we have \begin{align*}
    \chi_r(z)&=\tau_r\left((z-{\textstyle\sum}_{s=1}^kh_s^*t_s\upsilon_s(d))^{-1}h_r^*t_r\right)\\
    &=\frac{m_0}{m_r}\sum_{l\geq 0}z^{-(l+1)}\tau_0(({\textstyle\sum}_{s=1}^kh_s^*t_s\upsilon_s(z))^lh_r^*t_r).
\end{align*}
Applying \eqref{eq:d.law.H}, then we further have
\begin{align*}
    \chi_r(z)
    &=\sum_{l\geq 0}z^{-(l+1)}\frac{1}{m_r}\Tr (({\textstyle\sum}_{s=1}^kH_s^*T_s\upsilon_s(z))^lH_r^*T_r)\\
    &=\frac{1}{m_r}\Tr \left(\left(z\Id _{l_0}-{\textstyle\sum}_{s=1}^k\upsilon_s(z)H_s^*T_s\right)^{-1}H_r^*T_r\right).
\end{align*}   
}
Define $u:=\sum_{r,s=1}^kf_{rs}$. Note that $\nu$ and $u$ are identical to \cite{fan2017eigenvalue}, the steps leading to equation (4.29) in \cite{fan2017eigenvalue} still apply. As a result, define $D=\text{diag}(D_1\Id _{n_1},\ldots,D_k\Id _{n_k})$ for $D_r=m_r\chi_r(z)/n_r$, we have
\begin{align*}
    \upsilon_r(z)& = \frac{1}{n_r}\Tr _r\left((\Id _{n_+}-FD)^{-1}F\right).
\end{align*}
Setting $x_r=-zm_r/n_r\chi_r$, $y_r=-z^{-1}\upsilon_r$, we obtain
\begin{align*}
    x_r
    &=-\frac{1}{n_r}\Tr \left(\left(\Id _{l_0}+{\textstyle\sum}_{s=1}^ky_rH_s^*T_s\right)^{-1}H_r^*T_r\right),\\
    y_r& = -\frac{1}{n_r}\Tr _r\left((z\Id _{n_+}+FD(x))^{-1}F\right).
\end{align*}

By definition of the conditional expectation and the Cauchy transform in \cite{fan2017eigenvalue}, we can alternatively write $\chi_r(z)$ defined by \eqref{eq:chi.r} as  
\begin{align*}
    \chi_r(z)&:=\tau_r(\sum_{l\geq 0}\mathbf{F}^{\mathcal{H}}(z^{-1}(\omega z^{-1})^l)h_r^*t_r)\\
    &=\sum_{l\geq 0}z^{-(l+1)}\tau_r(\mathbf{F}^{\mathcal{H}}((\omega)^l)h_r^*t_r)=\frac{N}{m_r}\sum_{l=0}^{\infty}z^{-(l+1)}\tau(\omega^lh_r^*t_r)
\end{align*}
for $z\in\mathbb{D}$. Since $|\tau(\omega^lh_r^*t_r)|\leq \lVert\omega\rVert^l\lVert h_r\rVert\cdot\lVert t_r\rVert$, $\chi_r$ defines an analytic function on $\mathbb{D}$ such that $\chi_r(z)\sim (zm_r)^{-1}\Tr (H_r^*T_r)$ as $|z|\rightarrow \infty$.
Note that
\begin{align*}
    \frac{1}{m_r}\Tr ((z\Id -W)^{-1}H_r^*T_r)=\frac{N}{m_r}\sum_{l=0}^{\infty}z^{-(l+1)}N^{-1}\Tr (W^lH_r^TT_r).
\end{align*}
Applying Theorem 3.10 and adapting the proof of Corollary 3.11 in \cite{fan2017eigenvalue}, we have for $z\in\mathbb{D}$, 
\begin{align*}
    x_r-z \frac{1}{n_r}\Tr ((W-z\Id _{l_0})^{-1}H_r^*T_r)\rightarrow_{a.s.}0.
\end{align*}

\section{Additional Proofs}\label{sec:proofs}
\subsection{Proof of Lemma \ref{lem:deteqv}}\label{app:deteqv}
    Applying Theorem 1.2 in \cite{fan2017eigenvalue}, we have that for each $z\in\mathbb{C}^+$, there exists unique z-dependent values ${a}_n^{(1)},\ldots,{a}_n^{(k)}\in\mathbb{C}^+\cup\{0\}$, ${b}_n^{(1)},\ldots,{b}_n^{(k)}\in\overline{\mathbb{C}^+}$, and $\tilde{m}_n\in\mathbb{C}^+$, such that 
    \begin{align}\label{eq:stieltjes.fj}
        \begin{cases}
        \tilde{m}_n&=-\frac{1}{p}\Tr ((z\Id +\sum_{r=1}^kb_n^{(r)}\Sigma_r)^{-1})\\
            {a}_n^{(r)}&=-\frac{1}{I_r}\Tr ((z\Id +\sum_{s=1}^kb_n^{(s)}\Sigma_s)^{-1}\Sigma_r),\quad r=1,\ldots,k\\
            {b}_n^{(r)}&=-\frac{1}{I_r}\Tr _r([\Id _{I_+}+FD(a)]^{-1}F),\quad r=1,\ldots,k,
        \end{cases}
    \end{align}
    where $D(a) = \text{diag}(a_n^{(1)}\Id _{I_1},\ldots,a_n^{(k)}\Id _{I_k})\in\mathbb{C}^{I_+\times I_+}$. 
The function $\tilde{m}_n:\mathbb{C}^+\rightarrow\mathbb{C}^+$ defines the Stieltjes transform of a probability measure $\tilde{\mu}_n$ on $\mathbb{R}$ such that
\begin{align*}
    \tilde{m}_n(z)-m_n(z)\rightarrow_{a.s} 0,
\end{align*}
and $\tilde{\mu}_n-\mu_n\rightarrow0$ weakly almost surely.

Let $\tilde{a}_n^{(r)}=I_r/p{a}_n^{(r)}$, then we immediately have that
\begin{align*}
    \tilde{a}_n^{(r)}&=-\frac{1}{p}\Tr ((z\Id +\sum_{s=1}^kb_n^{(s)}\Sigma_s)^{-1}\Sigma_r).
\end{align*}

It remains to show that $b_n^{(r)}$ defined in \eqref{eq:stieltjes.fj} equals $b_n^{(r)}$ defined in \eqref{eq:b.n.r}. We apply the matrix identity $(AB +\Id)^{-1}A = A(BA + \Id)^{-1}$, setting $A = RL$ and $B = R^TD(a)$, and yielding
\begin{align}
    (\Id _{I_+}+FD(a))^{-1}F &= RL(\Id+R^TD(a) RL)^{-1}R^T.\label{eq:br.sherman.morrison}
\end{align}
From the definitions, $R^TD(a) R = \sum_r{a}_n^{(r)}R_r^TR_r= \sum_r\tilde{a}_n^{(r)}\tilde{R}_r^T\tilde{R}_r$, and so taking the $r$th block trace,
\begin{equation}
  \Tr_r [FD(a)+\Id]^{-1}F
  = \Tr[(L^{-1}+{\textstyle\sum}_{s}\tilde{a}_n^{(s)}\tilde{R}_s^T\tilde{R}_s)^{-1}R_r^TR_r],\label{eq:br.sherman.morrison.2}
\end{equation}
whence
\begin{equation*}
    b_n^{(r)}= -\frac{1}{p}\Tr ((L^{-1}+{\textstyle\sum}_{s}\tilde{a}_n^{(s)}\tilde{R}_s^T\tilde{R}_s)^{-1}\tilde{R}_r^T\tilde{R}_r).
\end{equation*}

   \subsection{Addtional justifications for \eqref{eq:bn.bn.tilde}}\label{app:bn.bn.tilde}
Recall that we can write $S_n$ and $\mathcal{S}_n$ as 
\begin{align*}
    S_n &=\alpha^TRLR^T\alpha^T,\quad  \mathcal{S}_n=R^T\alpha\alpha^TRL.
\end{align*}  
Define 
\begin{align*}
    A=\alpha^TRL ,\quad C=R^T\alpha,
  \end{align*}
so that
$$
S_n=AC,\quad \mathcal{S}_n=CA.
$$
By the Sherman–Morrison formula,
\begin{align*}
   &(z^{-1}\mathcal{S}_n-I)^{-1}=(-I+Cz^{-1}IA)^{-1}=-I-C(z\Id -AC)^{-1}A=-I+C(S_n-z\Id )^{-1}A,
\end{align*}
which implies that
\begin{align*}
    (\mathcal{S}_n-z\Id )^{-1}=-z^{-1}I+z^{-1}C(S_n-z\Id )^{-1}A.
\end{align*}
From \eqref{eq:f}, we have $F=\frac{1}{p}U^TBU=RLR^T$. It follows that
\begin{align*}
    \Tr ((\mathcal{S}_n-z\Id )^{-1}R_r^TT_r)&=\Tr (-z^{-1}R_r^TT_r)+z^{-1}\Tr _r(F\alpha(S_n-z\Id )^{-1}\alpha^TF)\\
    &=-z^{-1}\Tr (R_r^TR_rL)+z^{-1}\frac{1}{p^2}\Tr _r(U^TBY(S_n-z\Id )^{-1}Y^TBU)\\
    &=-z^{-1}\Tr (R_r^TR_rL)+z^{-1}\frac{I_r}{p^2}\Tr (U_r^TBY(S_n-z\Id )^{-1}Y^TBU_r).
\end{align*}
We conclude that
\begin{align*}
    & b_n^{(r)}-(-I_r^{-1})z\Tr ((\mathcal{S}_n-z\Id )^{-1}R_r^TR_rL)\\
    =&b_n^{(r)}-\left[-\Tr (R_r^TR_rL)/I_r+\frac{1}{p}\Tr ((S_n-z\Id )^{-1}\frac{1}{p}Y^TBU_rU_r^TBY)\right]\\
    =&\tilde{b}_n^{(r)}(z)-\frac{1}{p}\Tr ((S_n-z\Id )^{-1}S_n^{(r)})\\
    =&
    \tilde{b}_n^{(r)}(z)-m_{n,b}^{(r)}\rightarrow_{a.s.}0.
\end{align*}

\subsection{Proof of Lemma \ref{lem:singular}}\label{app:proof.singular}
For each $r=1,\ldots,k$, the measure $\mu_{n,a}^{(r)}$ defined in (\ref{eq:mu.n.a}) can be decomposed into its restriction to $\mathbb{R}^*$ and its atom at $0$:
\begin{align*}
    \mu_{n,a}^{(r)}=\frac{\kappa_n}{p}\mu_{n,a,\mathbb{R}^*}^{(r)}+\frac{p-\kappa_n}{p}d_{n,0}^{(r)}\delta_0,\quad \mu_{n,a,\mathbb{R}^*}^{(r)} =\frac{1}{\kappa_n}\sum_{i=1}^{p}d_{n,i}^{(r)}\delta_{\lambda_{n,i}}\mathbbm{1}[\lambda_{n,i}\neq0].
\end{align*}
Similarly, we can write 
\begin{align*}
    \tilde{\mu}_{n,a}^{(r)}=\frac{\kappa_n}{p}\tilde{\mu}_{n,a,\mathbb{R}^*}^{(r)}+\frac{p-\kappa_n}{p}\tilde{d}_{n,0}^{(r)}\delta_0,
\end{align*}
with the Stieltjes transform of $\tilde{\mu}_{n,a}^{(r)}$ given by
\begin{align*}
    \tilde{a}_{n}^{(r)}(z)=\frac{\kappa_n}{p}\tilde{a}_{n,\mathbb{R}^*}^{(r)}(z)-z^{-1}\frac{p-\kappa_n}{p}\tilde{d}_{n,0}^{(r)}.
\end{align*}
Combining the above with the relationship established in (\ref{eq:b.n.r}), we have for $r=1,\ldots,k$,
\begin{align*}
{b}_n^{(r)}(z)&=-\frac{1}{p}\Tr ((L^{-1}+{\textstyle{\sum}}_{s=1}^k\tilde{a}_n^{(s)}\tilde{R}_s^T\tilde{R}_s)^{-1}\tilde{R}_r^T\tilde{R}_r)\\
&=-z\frac{1}{p}\Tr \left([zL^{-1}+p^{-1}{\textstyle{\sum}}_{s=1}^k(z\kappa_n\tilde{a}_{n,\mathbb{R}^*}^{(r)}(z)-{(p-\kappa_n)}\tilde{d}_{n,0}^{(r)})\tilde{R}_s^T\tilde{R}_s]^{-1}\tilde{R}_r^T\tilde{R}_r\right).
\end{align*}

By construction of $F = R L R^T$ as a spectral decomposition, with $L$ being full rank, and $R \in \mathbb{R}^{I_+ \times l_0}$ with orthonormal columns, we have $R^TR=\sum_{r=1}^k R_r^TR_r=\Id _{l_0}$. By the definition of $d_{n,0}^{(r)}$ above, for each $r$, $d_{n,0}^{(r)}=0$ if and only if $\text{null}(S_n)\subseteq\text{null}(\Sigma_r)$. With each $\Sigma_r$ being full rank, then $d_{n,0}^{(r)}>0$ for all $r$, and ${\textstyle{\sum}}_{s=1}^k\tilde{d}_{n,0}^{(s)}R_s^TR_s$ is invertible, with ${\textstyle{\sum}}_{s=1}^k\tilde{d}_{n,0}^{(s)}x^TR_s^TR_sx\geq \min_r\tilde{d}_{n,0}^{(r)}\|x\|^2$ for all $x\in\mathbb{R}^{l_0}$. Consequently, together with $z\tilde{a}_{n,\mathbb{R}^*}^{(r)}(z)\rightarrow0$, we have 
\begin{align}\label{eq:b.n.r.0}
    \lim_{z\in\mathbb{C}^+\rightarrow0}{b}_n^{(r)}(z)=0.
\end{align}
Combined with the definition in (\ref{eq:tilde.b.n.r}), define $\check{b}_n^{(r)}(0):= \lim_{z\rightarrow0^+}\tilde{b}_n^{(r)}(z)=\Tr (R_r^TR_rL)/I_r$. Then
\begin{align*}
    \check{b}_n^{(r)}{}'(0):=\lim_{z\rightarrow0^+}\frac{\tilde{b}_n^{(r)}(z)-\check{b}_n^{(r)}(0)}{z}=\frac{1}{p-\kappa_n}\Tr (({\textstyle{\sum}}_{s=1}^k\tilde{d}_{n,0}^{(s)}\tilde{R}_s^T\tilde{R}_s)^{-1}\tilde{R}_r^T\tilde{R}_r).
\end{align*}

\section{The singular case}\label{app:singular}
Section \ref{subsec:singular} describes the treatment of the null space of $S_n$. We now consider the nonzero eigenvalues of $S_n$; the complete procedure is summarized in Algorithm \ref{alg:singular}. As discussed in Section \ref{subsec:singular}, the derivation of the deterministic equivalent oracle estimator given in \eqref{eq:deteqv.oracle} remains valid for the nonzero eigenvalues. Hence, for each $\lambda_{n,i}\neq 0$, the deterministic equivalent oracle estimator of $\tau_{n,i}$ is 
\begin{align*}
    \check{d}_{n,i}=\frac{\Im [\check{a}_n^{(1)}(\lambda_{n,i})]}{\Im [\check{m}_n(\lambda_{n,i})]}.
\end{align*}
Because $S_n$ has a nontrivial null space, estimation of $\check{a}_n^{(1)}(\lambda_{n,i})$ and $\check{m}_n(\lambda_{n,i})$ is based on the empirical spectral distribution restricted to the nonzero eigenvalues. Recall from Section \ref{subsec:singular} that
\begin{align*}
    \kappa_n = \text{rank}(S_n), \quad d_{n,0}^{(r)}
    = \frac{1}{p-\kappa_n}\sum_{i=1}^{p}
    d_{n,i}^{(r)}\mathbbm{1}\{\lambda_{n,i}=0\},
\end{align*}
and
\begin{align*}
    m_{n,b,\mathbb{R}^*}^{(r)}(z)=\frac{1}{\kappa_n}\sum_{i=1}^{p}\frac{t_{n,i}^{(r)}}{\lambda_{n,i}-z}\mathbbm{1}[\lambda_{n,i}\neq0].
\end{align*}
Similarly, define
\begin{gather*}
    t_{n,0}^{(r)}
    = \frac{1}{p-\kappa_n}\sum_{i=1}^{p}
    t_{n,i}^{(r)}\mathbbm{1}\{\lambda_{n,i}=0\},\\
    m_{n,\mathbb{R}^*}=\frac{1}{\kappa_n}\sum_{i=1}^{p}\frac{1}{\lambda_{n,i}-z}\mathbbm{1}[\lambda_{n,i}\neq0],\quad m_{n,a,\mathbb{R}^*}^{(r)}(z)=\frac{1}{\kappa_n}\sum_{i=1}^{p}\frac{d_{n,i}^{(r)}}{\lambda_{n,i}-z}\mathbbm{1}[\lambda_{n,i}\neq0],
\end{gather*}
so that
\begin{align}
    m_{n}(z) &= \frac{\kappa_n}{p}m_{n,\mathbb{R}^*}(z)+z^{-1}\left(\frac{\kappa_n}{p}-1\right),\label{eq:m.n.decompose}\\
    m_{n,a}^{(r)}(z) &= \frac{\kappa_n}{p}m_{n,a,\mathbb{R}^*}^{(r)}(z)+z^{-1}\left(\frac{\kappa_n}{p}-1\right){d}_{n,0}^{(r)},\nonumber\\
    m_{n,b}^{(r)}(z) &= \frac{\kappa_n}{p}m_{n,b,\mathbb{R}^*}^{(r)}(z)+z^{-1}\left(\frac{\kappa_n}{p}-1\right)t_{n,0}^{(r)}.\label{eq:m.n.b.decompose}
\end{align}
Applying kernel smoothing to $m_{n,\mathbb{R}^*}$ and $m_{n,b,\mathbb{R}^*}^{(r)}$, define
\begin{align*}
    \underline{\hat{f}}_n(x) &= \frac{1}{\kappa_n}\sum_{i=1}^p\frac{1}{h_{n,i}}\kappa^E\left(\frac{x-\lambda_{n,i}}{h_{n,i}}\right)\mathbbm{1}[\lambda_{n,i}\neq0],\\
    \underline{\hat{g}}_n^{(r)}(x) &= \frac{1}{\kappa_n}\sum_{i=1}^p\frac{t_{n,i}^{(r)}}{h_{n,i}}\kappa^E\left(\frac{x-\lambda_{n,i}}{h_{n,i}}\right)\mathbbm{1}[\lambda_{n,i}\neq0].
\end{align*}
We then compute $\mathcal{H}_{\underline{\hat{f}}_n}$ and $\mathcal{H}_{\underline{\hat{g}}_n^{(r)}}$ analogously to the full-rank case, and set
\begin{align*}
    {\hat{m}}_{n,\mathbb{R}^*}(x)&=\pi\mathcal{H}_{\underline{\hat{f}}_n}(x)+i\pi \underline{\hat{f}}_n(x),\quad{\hat{b}}_{n,\mathbb{R}^*}^{(r)}(x)=\pi\mathcal{H}_{\underline{\hat{g}}_n^{(r)}}(x)+i\pi \underline{\hat{g}}_n^{(r)}(x).
\end{align*}
Together with (\ref{eq:m.n.decompose}) and (\ref{eq:m.n.b.decompose}), set the bona fide estimators of $\check{m}_{n}(z)$ and $\check{b}_{n}^{(r)}(z)$ as
\begin{align*}
    \hat{m}_{n}(x)&=\frac{\kappa_n}{p}\hat{m}_{n,\mathbb{R}^*}(x)+x^{-1}\frac{\kappa_n-p}{p},\\
    \hat{b}_{n}^{(r)}(x)&=\frac{\kappa_n}{p}\hat{b}_{n,\mathbb{R}^*}^{(r)}(x)+x^{-1}\frac{\kappa_n-p}{p}t_{n,0}^{(r)}.
\end{align*}
Similarly as in the full rank case, for each $x\neq0$, set $(\hat{a}_{n}^{(1)}(x),\ldots,\hat{a}_{n}^{(k)}(x))$ as the solution to the system of equations given in (\ref{eq:solve.bonafide}):
\begin{align*}
\begin{cases}
    \hat{b}_{n}^{(r)} &= -\frac{1}{p}\Tr \left((L^{-1}+\sum_{s=1}^k\hat{a}_{n}^{(s)}\tilde{R}_s^T\tilde{R}_s)^{-1}\tilde{R}_r^T\tilde{R}_r\right)+\frac{1}{p}\Tr (\tilde{R}_r^T\tilde{R}_rL).
\end{cases}
\end{align*}
Finally, for each $\lambda_{n,i}\neq 0$, we estimate $\tau_{n,i}$, the $i$th largest eigenvalue of $\Sigma_1$, by
\begin{align*}
    \hat{d}_{n,i} &= \frac{\Im \hat{a}_n^{(1)}(\lambda_{n,i})}{\Im \hat{m}_n(\lambda_{n,i})}.
\end{align*}

\begin{algorithm}[!htbp]
    \caption{Nonparametric Shrinkage Estimation for $\Sigma_1$ ($S_n$ singular)}\label{alg:singular}
    \textbf{Input:}
    \begin{algorithmic} 
    \State Same as Algorithm \ref{alg:full}.
    \end{algorithmic}
    \textbf{Procedure:}
    \begin{algorithmic}[1] 
        \State Steps 1--3 are the same as Algorithm \ref{alg:full}.
        \setcounter{ALG@line}{3} 
    \State Obtain $\kappa_n=\text{rank}(S_n)$, and compute the kernel smoothed p.d.f of the ESD of $S_n$ and its Hilbert transform as
    \[
        \underline{\hat{f}}_n(x) = \tfrac{1}{\kappa_n}{\textstyle\sum}_{i=1}^p \tfrac{1}{h_{n,i}}\kappa^E\left(\tfrac{x-\lambda_{n,i}}{h_{n,i}}\right)\mathbbm{1}[\lambda_{n,i}\neq0], \quad 
    \mathcal{H}_{\underline{\hat{f}}_n}(x) = \tfrac{1}{\kappa_n} {\textstyle\sum}_{i=1}^p \tfrac{1}{h_{n,i}} \mathcal{H}_{\kappa^E} \Bigl( \tfrac{x-\lambda_{n,i}}{h_{n,i}} \Bigr)\mathbbm{1}[\lambda_{n,i}\neq0],
\]
     where $\kappa^E$ and $\mathcal{H}_{\kappa^E}$ are respectively the Epanechnikov kernel and its Hilbert transform as in (\ref{eq:kappa.E}). 
     \State Obtain the kernel smoothed companion Stiletjes transform
     ${\hat{m}}_{n,\mathbb{R}^*}(x)=\pi\mathcal{H}_{\underline{\hat{f}}_n}(x)+i\pi \underline{\hat{f}}_n(x)$, and the Stieltjes transform $\hat{m}_n(x)=(\kappa_n/p){\hat{m}}_{n,\mathbb{R}^*}(x)+x^{-1}(\kappa_n/p-1)$.
    \For{\texttt{$r \gets 1$ to $k$}}
    \State Construct matrix $S_n^{(r)} = \frac{1}{p}Y^TBU_rU_r^TBY$, and compute weights $t_{n,i}^{(r)}=u_{n,i}^TS_n^{(r)}u_{n,i}$ for all $i$ such that $\lambda_{n,i}\neq0$, and $t_{n,0}^{(r)}=\frac{1}{p-\kappa_n}\sum_{i=1}^p t_{n,i}^{(r)}\mathbbm{1}[\lambda_{n,i}=0]$.
    \State Compute the weighted measure and its Hilbert transform   
    \[
        \underline{\hat{g}}_n^{(r)}(x) = \tfrac{1}{\kappa_n} {\textstyle\sum}_{i=1}^p \tfrac{t_{n,i}^{(r)}}{h_{n,i}} \kappa^E \Bigl( \tfrac{x-\lambda_{n,i}}{h_{n,i}} \Bigr)\mathbbm{1}[\lambda_{n,i}\neq0], \quad 
        \mathcal{H}_{\underline{\hat{g}}_n^{(r)}}(x) = \tfrac{1}{\kappa_n} {\textstyle\sum}_{i=1}^p \tfrac{t_{n,i}^{(r)}}{h_{n,i}} \mathcal{H}_{\kappa^E} \Bigl( \tfrac{x-\lambda_{n,i}}{h_{n,i}} \Bigr)\mathbbm{1}[\lambda_{n,i}\neq0].
    \]
   \State Obtain the Stieltjes transform of the weighted measure
    ${\hat{b}}_{n,\mathbb{R}^*}^{(r)}(x)=\pi\mathcal{H}_{\underline{\hat{g}}_n^{(r)}}(x)+i\pi \underline{\hat{g}}_n^{(r)}(x)$, and the Stieltjes transform $\hat{b}_{n}^{(r)}(x)=(\kappa_n/p){\hat{b}}_{n,\mathbb{R}^*}^{(r)}(x)+x^{-1}(\kappa_n/p-1)t_{n,0}^{(r)}$.
    \State Obtain the estimated derivative $\hat{b}_{n}^{(r)}{}'(0)=\frac{1}{p}\sum_{i=1}^{p}\frac{t_{n,i}^{(r)}}{\lambda_{n,i}^2}\mathbbm{1}[\lambda_{n,i}\neq0].$
    \EndFor
    \State For each $\lambda_{n,i}\neq0$, set $(\hat{a}_{n}^{(1)}(\lambda_{n,i}),\ldots,\hat{a}_{n}^{(k)}(\lambda_{n,i}))$ as the solution to the system of equations
    \begin{align*}
    \begin{cases}
            \hat{b}_{n}^{(r)}(\lambda_{n,i}) &= -\frac{1}{p}\Tr \left((L^{-1}+\sum_{s=1}^k\hat{a}_{n}^{(s)}(\lambda_{n,i})\tilde{R}_s^T\tilde{R}_s)^{-1}\tilde{R}_r^T\tilde{R}_r\right)+\frac{1}{p}\Tr (\tilde{R}_r^T\tilde{R}_rL).
    \end{cases}
\end{align*}
Set $(\hat{d}_{n,0}^{(1)},\ldots,\hat{d}_{n,k}^{(k)})$ as the solutions to the system of equations
\begin{align}
    \begin{cases}
      \hat{b}_{n}^{(r)}{}'(0)=\frac{1}{p-\kappa_n}\Tr (({\textstyle{\sum}}_{s=1}^k\hat{d}_{n,0}^{(s)}\tilde{R}_s^T\tilde{R}_s)^{-1}\tilde{R}_r^T\tilde{R}_r).
    \end{cases}
\end{align}
    \State Denote the shrinkage estimator for the $i$-th largest eigenvalue of $\Sigma_1$ by $\hat{d}_{n,i}$. Set $\hat{d}_n=(\hat{d}_{n,1},\ldots,\hat{d}_{n,p})$ with $\hat{d}_{n,i}=\frac{\Im [\hat{a}_{n}^{(1)}(\lambda_{n,i})]}{\Im [{\hat{m}}_n(\lambda_{n,i})]}$ if $\lambda_{n,i}\neq0$, and $\hat{d}_{n,i}=\hat{d}_{n,0}$ if $\lambda_{n,i}=0$.
    
   \State (optionally) Perform isotonic regression on $\hat{d}_n$ as described in Section \ref{sec:isotonic}.
    \end{algorithmic}
    \textbf{Output:}
    \begin{algorithmic}
     \State Nonparametric shrinkage estimator $\hat{\Sigma}_1=\mathbb{U}_n\text{diag}(\hat{d}_n)\mathbb{U}_n^T$
    \end{algorithmic}
\end{algorithm}

\section{Specialization to the single-component setting}
When there is a single variance component, $S_n$ reduces to either a Wishart matrix or a separable covariance matrix. In the Wishart setting, \cite{Ledoit2011} derives the deterministic equivalent oracle shrinkage formula, and \cite{ledoit2020analytical} develops a bona fide nonparametric estimator. For separable covariance matrices, \cite{oriol2024asymptotic} derives the corresponding deterministic equivalent oracle formula. We now specialize our results to this one-component setting.
\subsection{$S_n$ full rank}
The deterministic equivalent oracle is a direct generalization of the Wishart case \cite{Ledoit2011}, and the resulting formula agrees with that of \cite{oriol2024asymptotic}. We therefore focus on the bona fide nonparametric estimator, which has previously been studied only for Wishart matrices. In particular, we specialize our estimator to the Wishart setting and compare it with the method of \cite{ledoit2020analytical}.

When $k=1$, consider the following setup:
\begin{align*}
    Y=U_1X\Sigma^{1/2},\quad U_1=\Id _n,\quad X\in\mathbb{R}^{n\times p}, \quad \Sigma\in\mathbb{R}^{p\times p},
\end{align*}
where $X$ has i.i.d. standard Gaussian entries and $\Sigma$ is positive semidefinite. Let $B=(p/n)\Id_n$, and consider the sum-of-squares matrix
\begin{align*}
    S_n=\frac{1}{p}Y^TBY=\frac{1}{n}\Sigma^{1/2}X^TX\Sigma^{1/2}.
\end{align*}
Let $U=\sqrt{n}U_1$. Then $F=\frac{1}{p}U^TBU=\Id_n$, $R=L=\Id_n$, and $\tilde{R}=\sqrt{p/n}\Id_n$. By Lemma \ref{lem:deteqv}, the deterministic equivalent Stieltjes transform of $S_n$ is determined by
\begin{align}
\tilde{m}_n&=-\frac{1}{p}\Tr ((z\Id +b\Sigma)^{-1}),\nonumber\\
    \tilde{a}_n&=-\frac{1}{p}\Tr ((z\Id +b\Sigma)^{-1}\Sigma),\nonumber\\
    b_n&=-\frac{1}{n}\Tr ((\Id _{n}+(p/n)\tilde{a}\Id _n)^{-1})=-\frac{1}{1+(p/n)\tilde{a}_n}.\label{eq:wishart.deteqv.system}
\end{align}

The corresponding oracle estimator is
\begin{align*}
    d_{n,i}=
    \frac{\Im [\check{a}_n(\lambda_i+i\eta)]}{\Im [\check{m}_n(\lambda_i+i\eta)]}.
\end{align*}
By Lemma \ref{lem:inversion.borel}, define 
\begin{align*}
    \tilde{b}_n=b_n+1, \quad S_n^{(1)}=\frac{1}{p}Y^TBU_1U_1^TBY=\frac{p}{n}S_n.
\end{align*}
Then
\begin{align*}
    \tilde{b}_n-m_{n,b}=\tilde{b}_n-\frac{1}{n}\Tr ((S_n-z\Id )^{-1}S_n)\rightarrow_{a.s.}0.
\end{align*}
Note that by definition,
\begin{align*}
    t_{n,i}^{(1)}=u_{n,i}^TS_n^{(1)}u_{n,i}=\frac{p}{n}\lambda_{n,i},
\end{align*}
then
\begin{align*}
    \hat{f}_n(x) = \frac{1}{p}\sum_{i=1}^p\frac{1}{h_{n,i}}k\left(\frac{x-\lambda_{n,i}}{h_{n,i}}\right),\\
     \hat{g}_n(x) = \frac{1}{p}\sum_{i=1}^p\frac{p/n\cdot\lambda_{n,i}}{h_{n,i}}k\left(\frac{x-\lambda_{n,i}}{h_{n,i}}\right).
\end{align*}
As a result,
\begin{align*}
\hat{m}_n(\lambda_i)&=\pi\mathcal{H}_{\hat{f}_n}(\lambda_i)+i\pi\hat{f}_n(\lambda_i),\\
   \hat{b}_n(\lambda_i)&= \pi\mathcal{H}_{\hat{g}_n}(\lambda_i)+i\pi \hat{g}_n(\lambda_i).
\end{align*}
Combining the formula $\tilde{b}_n=b_n+1$ with \eqref{eq:wishart.deteqv.system}, we obtain
\begin{align*}
\hat{a}_n(\lambda_i)&= -\frac{n}{p}\left(1+\frac{1}{\hat{b}_n(\lambda_i)-1}\right)= -\frac{n}{p}\left(1+\frac{-1+\pi\mathcal{H}_{\hat{g}_n}(\lambda_i)-i\pi\hat{g}_n(\lambda_i)}{|-1+\pi\mathcal{H}_{\hat{g}_n}(\lambda_i)|^2+|\pi\hat{g}_n(\lambda_i)|^2}\right).
\end{align*}
Collecting the terms, we conclude that the bona fide nonparametric estimator of $\tau_{n,i}$ is given by
\begin{align*}
\Im [\hat{a}_n(\lambda_i)]&=\frac{n}{p}\frac{\pi\hat{g}_n(\lambda_i)}{|-1+\pi\mathcal{H}_{\hat{g}_n}(\lambda_i)|^2+|\pi\hat{g}_n(\lambda_i)|^2},\quad \Im [\hat{m}_n(\lambda_i)]=\pi\hat{f}_n(\lambda_i),\\
    \hat{d}_{n,i}=
    \frac{\Im [\hat{a}_n(\lambda_i)]}{\Im [\hat{m}_n(\lambda_i)]}&= \frac{n}{p}\frac{\hat{g}_n(\lambda_i)}{\hat{f}_n(\lambda_i)}\frac{1}{|1-\pi\mathcal{H}_{\hat{g}_n}(\lambda_i)|^2+|\pi\hat{g}_n(\lambda_i)|^2}.
\end{align*}

\begin{figure}[htb]
    \centering
    \includegraphics[width=\linewidth]{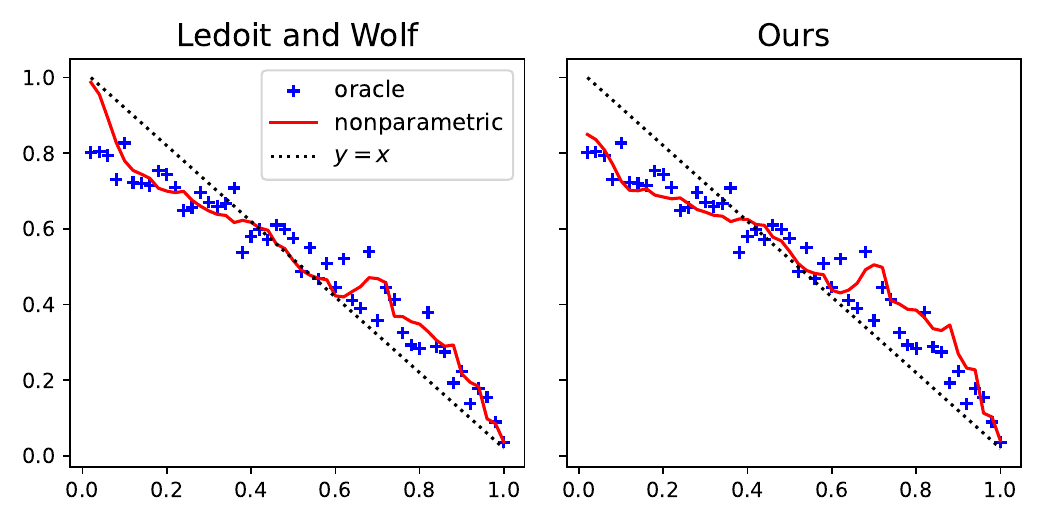}
    \caption{Oracle and bona fide nonparametric estimators under the Wishart setting.}
    \label{fig:sim.wishart}
\end{figure}

We compare the performance of the proposed estimator with the kernel estimator of \cite[Section 4.7]{ledoit2020analytical}, which is given by
\begin{align*}
    \hat{d}_{n,i}^{\text{LW}}  = \frac{\lambda_i}{[1-\hat{c}_n-\hat{c}_n\lambda_i  \pi H_{\hat{f}_n}(\lambda_i)]^2+[\hat{c}_n\lambda_i  \pi\hat{f}_n(\lambda_i)]^2},
\end{align*}
where $\hat{c}_n=\frac{p}{n}$. Note that in the Wishart setting considered here, it is not necessary to approximate $\hat{b}_n$ using kernel smoothing, since 
\begin{align*}
    m_{n,b}(z) = \frac{1}{n}\Tr ((S_n-z\Id )^{-1}S_n) = \frac{p}{n}\frac{1}{p}\left(p+z\Tr ((S_n-z\Id )^{-1}) \right)=\frac{p}{n}+z\frac{p}{n}{m}_n(z).
\end{align*}
It is therefore sufficient to take
\begin{align*}
    {\hat{b}_n}^{\text{LW}}(\lambda_i) =\frac{p}{n}+\lambda_i\frac{p}{n}\hat{m}_n(\lambda_i)=\hat{c}_n+\hat{c}_n\lambda_i\pi\mathcal{H}_{\hat{f}_n}(\lambda_i)+i\hat{c}_n\lambda_i\pi\hat{f}_n(\lambda_i),
\end{align*}
and 
\begin{align*}
    {\hat{a}}_n^{\text{LW}}(\lambda_i)&= -\frac{n}{p}\left(1+\frac{1}{\hat{b}_n(\lambda_i)-1}\right)= -\frac{n}{p}\left(1+\frac{-1+\hat{c}_n+\hat{c}_n\lambda_i\pi\mathcal{H}_{\hat{f}_n}(\lambda_i)-i\hat{c}_n\lambda_i\pi\hat{f}_n(\lambda_i)}{|1-\hat{c}_n-\hat{c}_n\lambda_i\pi\mathcal{H}_{\hat{f}_n}(\lambda_i)|^2+|\hat{c}_n\lambda_i\pi\hat{f}_n(\lambda_i)|^2}\right),
    \end{align*}
yielding
\begin{align*}
    \frac{\Im [{\hat{a}}_n^{\text{LW}}(\lambda_i)]}{\Im [{\hat{m}}_n(\lambda_i)]} = \frac{\Im [{\hat{a}}_n^{\text{LW}}(\lambda_i)]}{\pi\hat{f}_n(\lambda_i)}=\hat{d}_{n,i}^{\text{LW}}.
\end{align*}
We set $p=50$ and $n=100$, and let the eigenvalues of $\Sigma$ decay linearly, with $\tau_i=i/p$ for the $i$th eigenvalue. Figure \ref{fig:sim.wishart} compares the proposed estimator with the kernel estimator of \cite{ledoit2020analytical}. Both estimators perform well and exhibit only minor differences.

\subsection{$S_n$ singular}\label{sec:singular.separable}
As discussed in Section~\ref{subsec:singular}, when $S_n$ is singular, its nonzero eigenvalues can be treated analogously to the full-rank case, whereas its null space requires separate consideration. In particular, new arguments are needed both to derive the deterministic equivalent oracle and to construct the bona fide nonparametric estimator. This section focuses on the treatment of the null space of $S_n$. We first specialize our deterministic equivalent oracle to the singular setting and show that it recovers the formulas of \cite{Ledoit2011} for Wishart matrices and \cite{oriol2024asymptotic} for separable covariance matrices. We then specialize our bona fide nonparametric estimator to the Wishart setting and compare it with the method of \cite{ledoit2020analytical}.

When $k=1$,
\begin{align*}
    S_n=Y^TBY=\Sigma^{1/2}X^TU_1^TBU_1X\Sigma^{1/2}.
\end{align*}
Let $T=nU_1^TBU_1$, then we recover the separable covariance matrix
\begin{align*}
    S_n=\frac{1}{n}\Sigma^{1/2}X^TTX\Sigma^{1/2}.
\end{align*}
Suppose that $p>n$ and that $T$ is full rank. By Gaussianity and rotational invariance, we may assume without loss of generality that $T$ is diagonal. Thus we take $L=T$, $R=\Id_n$, $\widetilde R=\sqrt{p/n}\Id_n$, and $\kappa_n=n$. The system \eqref{eq:singular.system} then reduces to
\begin{align*}
    \check{b}_n{}'(0)=\frac{1}{p-n}\Tr ((\check{d}_{n,0}\tilde{R}^T\tilde{R})^{-1}\tilde{R}^T\tilde{R})=\frac{n}{p-n}\frac{1}{\check{d}_{n,0}}.
\end{align*}
Let $\hat{c}_n=p/n>1$, then the deterministic equivalent oracle estimator corresponding to the null space of $S_n$ simplifies to
\begin{align*}
    \check{d}_{n,0}=\frac{1}{(\hat{c}_n-1)\check{b}_n{}'(0)}.
\end{align*}
\paragraph{Wishart matrix}
From \eqref{eq:wishart.deteqv.system}, we have that ${b}_n=-1+p/n+z(p/n)m_n$. Then
\begin{align*}
   \frac{b_n(z)}{z} = -z^{-1}(1-p/n)+(p/n)m_n(z)=\tilde{m}_{n,\mathbb{R}^*}(z),\quad \lim_{z\in\mathbb{C}^+\rightarrow0}\frac{b_n(z)}{z}=\check{m}_{n,\mathbb{R}^*}(z).
\end{align*}
where $\check{m}_{n,\mathbb{R}^*}(0)$ is denoted ${\check{m}}_{\underline{F}}(0)$ in \cite{Ledoit2011}, and $\hat{c}_n$ corresponds to $\gamma^{-1}$ in their notation. Thus our formula for $\check{d}_{n,0}$ matches that in \cite[Theorem 1.4]{Ledoit2011}.

Note that by definition,
\begin{align*}
    t_{n,i}^{(1)}=u_{n,i}^TS_n^{(1)}u_{n,i}=\frac{p}{n}\lambda_{n,i},
\end{align*}
then \eqref{eq:singular.b.n.r.prime} simplifies to
\begin{align*}
    \hat{b}_{n}{}'(0)=\frac{1}{n}\sum_{i=1}^{p}\frac{1}{\lambda_{n,i}}\mathbbm{1}[\lambda_{n,i}\neq0],
\end{align*}
and the system \eqref{eq:singular.system} simplifies to
\begin{align*}
    \hat{d}_{n,0}=\frac{1}{\frac{p-n}{p}\hat{b}_{n}{}'(0)}=\frac{1}{\frac{p-n}{p}\frac{1}{n}\sum_{i=1}^{p}\frac{1}{\lambda_{n,i}}\mathbbm{1}[\lambda_{n,i}\neq0]}.
\end{align*}
From equations (C.3) and (C.8) in \cite{ledoit2020analytical}, we have
\begin{align*}
    \hat{d}_{n,0}^{\text{LW}}=\frac{1}{C_n\frac{p-n}{p}\frac{1}{n}\sum_{i=1}^{p}\frac{1}{\lambda_{n,i}}\mathbbm{1}[\lambda_{n,i}\neq0]},
\end{align*}
where $C_n$ is a constant computed from the global bandwidth $h_n=n^{-1/3}$, and satisfies $C_n\to 1$ as $n\to\infty$. In summary, our bona fide estimator is asymptotically equivalent to the bona fide estimator derived in \cite{ledoit2020analytical}.

\paragraph{Separable covariance matrix}
Comparing Theorem~2.1 of \cite{oriol2024asymptotic} with Lemma~\ref{lem:deteqv}, we identify their function $X(z)$ as
\begin{align*}
    X(z)=\frac{b_n(z)}{z}.
\end{align*}
As a result,
\begin{align*}
    \check{X}(0)=\lim_{z\in\mathbb{C}^+\rightarrow0}\frac{b_n(z)}{z}=\check{b}_n{}'(0).
\end{align*}
It follows that
\begin{align*}
    \check{d}_{n,0}=\frac{1}{(\hat{c}_n-1)\check{X}(0)}.
\end{align*}
This is precisely the formula in \cite[Theorem~2.4]{oriol2024asymptotic}, with $\hat c_n$ corresponding to $c$ in their notation.

\section{Simulations under the full sib half sib model}\label{app:halfsib}

Consider an experimental design in which each of the $I_1$ sires is mated with $M$ dams, yielding a total of  $I_2=I_1M$ dams. The $m$-th dam mated with the $i$-th sire produces $J_{im}$ offspring, resulting in a total of $n=\sum_{i=1}^{I_1}\sum_{m=1}^{M}J_{im}$ offspring. Let  $J_i=\sum_{m=1}^{M}J_{im}$ denote the total number of offspring from sire $i$. For each offspring, we measure $p$ phenotypic traits. The observed trait matrix $Y\in\mathbb{R}^{n\times p}$ is assumed to follow the model
\begin{align}\label{eq:Y.twub}
    Y=X\beta+U_1\alpha_1+U_2\alpha_2+U_3\alpha_3, \quad \alpha_r \sim N(0, \Id _{I_r} \otimes \Sigma_r),
\end{align}
where $X=\mathbbm{1}_n$ is the fixed effects design matrix, $\beta=\mu^T$ is the fixed effect, 
$$
U_1=\text{diag}(\mathbbm{1}_{J_1},\ldots,\mathbbm{1}_{J_{I_1}})\in\mathbb{R}^{n\times {I_1}}
$$
is the membership matrix that assigns each offspring to the corresponding sire, 
$$
U_2=\text{diag}(\mathbbm{1}_{J_{11}},\ldots,\mathbbm{1}_{J_{1M}},\ldots,\mathbbm{1}_{J_{I_11}},\ldots,\mathbbm{1}_{J_{I_1M}})\in\mathbb{R}^{n\times {I_2}}
$$
is the membership matrix that assigns each offspring to the corresponding dam, $U_3=\Id _n$ is the membership matrix that assigns each individual to the corresponding individual effect, and each $\alpha_r\in\mathbb{R}^{I_r\times p}$ represents the $r$-th level random effect.

Denote by $S_n^{(r)} = p^{-1}Y^{T} B_r Y$ the MANOVA estimator of $\Sigma_r$. Let $\pi_1$, $\pi_2$, and $\pi_3$ be the orthogonal projections onto $\mathrm{col}(U_1)\ominus \mathrm{col}(\mathbbm{1}_n)$, $\mathrm{col}(U_2)\ominus \mathrm{col}(U_1)$, and $\mathrm{col}(U_3)\ominus \mathrm{col}(U_2)$, respectively. Extending the formulae in Appendix~F.2 of \cite{searle2009variance}, the coefficient matrices satisfy
\begin{align*}
    B_3 &= \frac{p\pi_3}{n-I_2},\qquad 
    B_2 = \frac{p\pi_2-(I_2-I_1)B_3}{n-l_1},\qquad 
    l_1=\sum_{i=1}^{I_1} J_i^{-1}\sum_{m=1}^M J_{im}^2,\\
    B_1 &= \frac{p\pi_1-(l_1-l_2)B_2-(I_1-1)B_3}{n-l_3},\qquad 
    l_2=\sum_{i=1}^{I_1} n^{-1}\sum_{m=1}^M J_{im}^2,\qquad 
    l_3=n^{-1}\sum_{i=1}^{I_1} J_i^2.
\end{align*}
Let $Z$ denote an arbitrary $n' \times m'$ matrix whose entries are i.i.d.\ standard normal, and let $B$ denote a fixed real symmetric matrix. Then $\mathbb{E}[Z^{T} B Z] = \mathrm{tr}(B)\,\Id_{m'}$. Applying this identity with the parametrization in \eqref{eq:Y.twub}, we have that the MANOVA estimators are unbiased, with
\begin{align*}
    \mathbb{E}(S_n^{(r)}) = \Sigma_r,\qquad r=1,2,3.
\end{align*}

\subsection{Simulations}

We consider an experiment design with $I_1=100$ sires, each mated with $M=2$ dams. For each offspring, we observe $p=50$ phenotypic traits. The number of offspring $\{J_{im}\}$ produced by the $m$th dam mated with the $i$th sire is modeled as an i.i.d.\ random variable taking values $1$ or $2$ with equal probability, $J_{im}\sim\mathrm{Unif}\{1,2\}$. We set $\Sigma_2=\Sigma_3=I_p$, and study two scenarios for the spectrum of $\Sigma_1$:
\begin{enumerate}
    \item \textbf{Linear decay.} The eigenvalues satisfy $\tau_i=\tau_{\max} (1-(i-1)/p)$, $i=1,\ldots,p$.
    \item \textbf{Step decay.} The eigenvalues satisfy $\tau_i=\tau_{\max}$ for $i=1,\ldots,p/2$, and $\tau_i=0$ for $i=p/2+1,\ldots,p$.
\end{enumerate}

\begin{figure}[htb]
    \centering
    \includegraphics[width=0.8\linewidth]{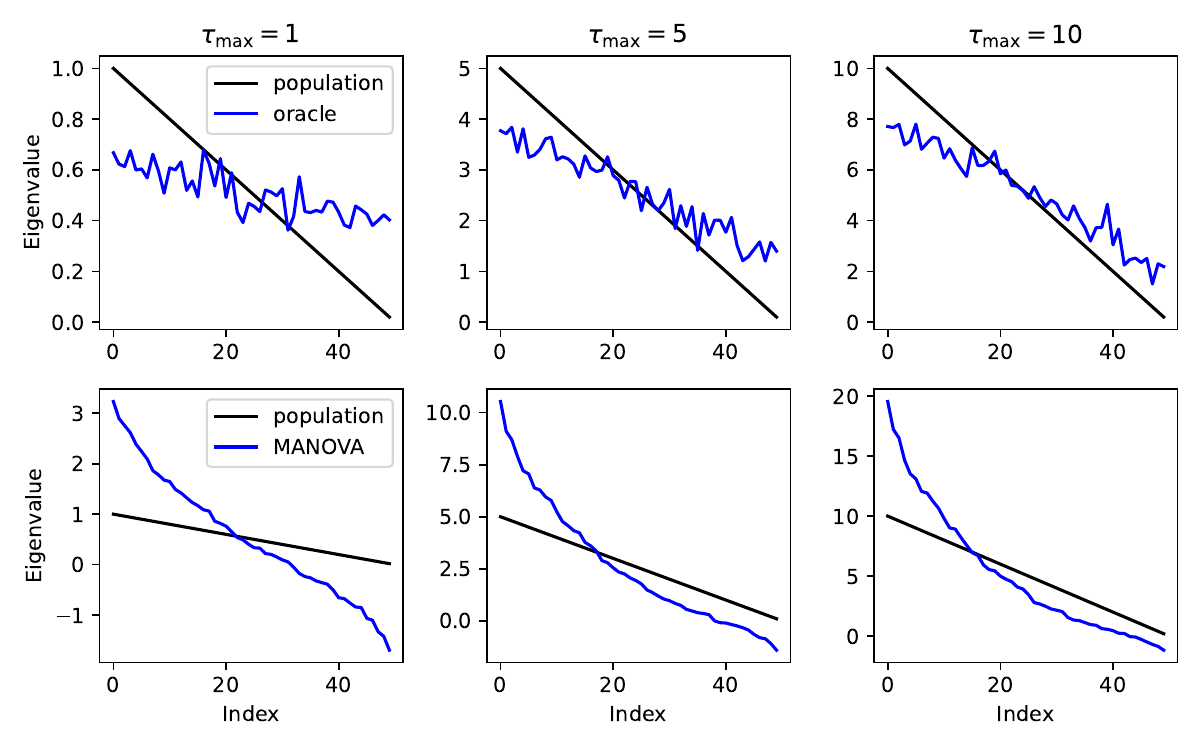}
    \caption{Linear decay: true eigenvalues versus oracle nonparametric and MANOVA estimates.}
    \label{fig:twub.oracle.linear}
\end{figure}

\begin{figure}[htb]
    \centering
    \includegraphics[width=0.8\linewidth]{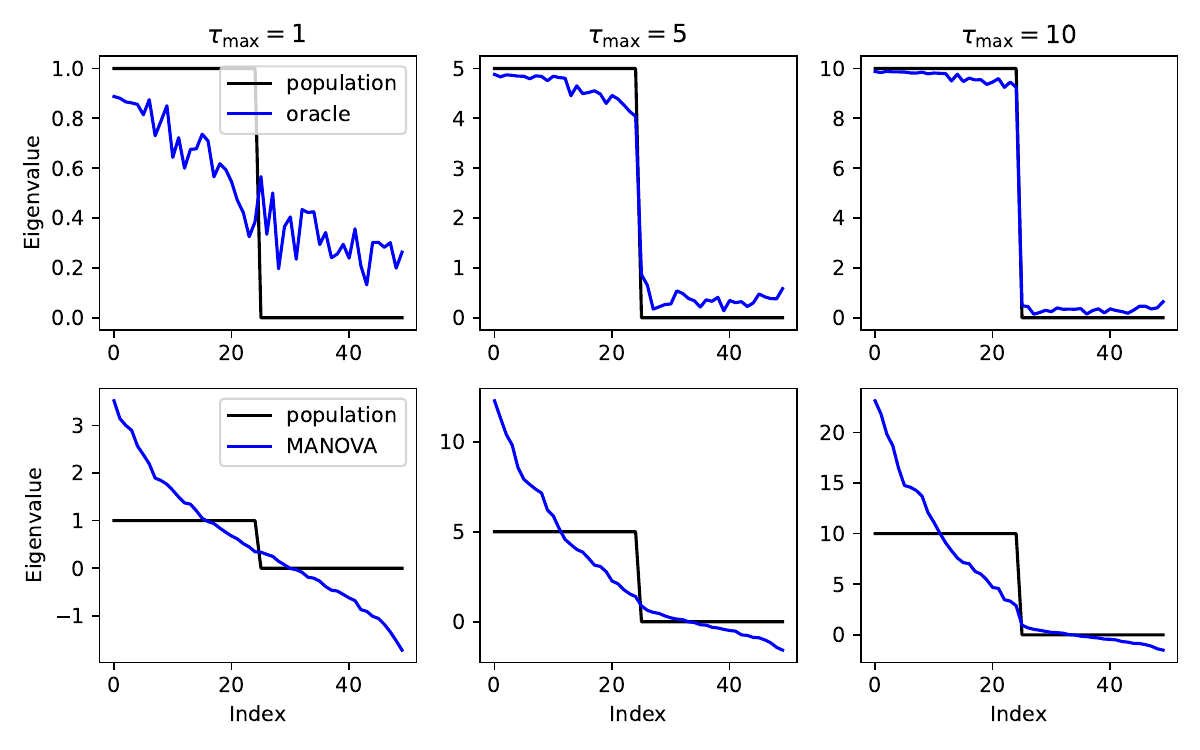}
    \caption{Step decay: true eigenvalues versus oracle nonparametric and MANOVA estimates.}
    \label{fig:twub.oracle.step}
\end{figure}

\paragraph{Oracle estimator.}
Figures~\ref{fig:twub.oracle.linear} and~\ref{fig:twub.oracle.step} compare the true eigenvalues of $\Sigma_1$ with the oracle nonparametric estimates $d_{n,i}^{\text{or}}$ and with the MANOVA eigenvalue estimates. As the eigenvalue scale increases, the oracle nonparametric estimates track the true eigenvalues more closely. By contrast, the MANOVA eigenvalue estimates exhibit significant bias compared to the population spectrum. The contrast is especially clear under step decay: the oracle estimates reproduce the pattern of large eigenvalues near $\tau_{\max}$, whereas the MANOVA estimates exhibit an increasing slope in the upper part of the spectrum.

\begin{figure}[htb]
    \centering
    \begin{subfigure}[t]{\textwidth}
        \centering
        \includegraphics[width=0.8\linewidth]{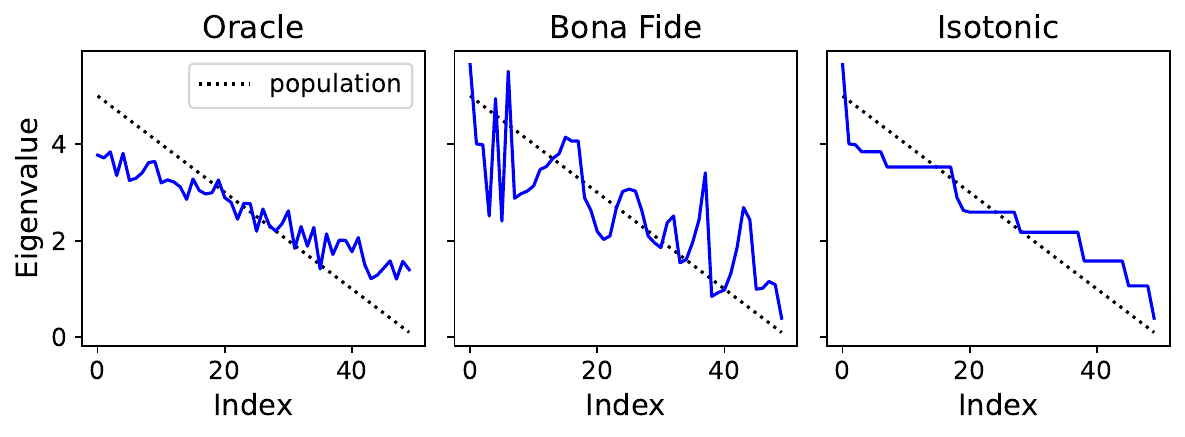}
        \caption{Linear decay.}
    \end{subfigure}
    \begin{subfigure}[t]{\textwidth}
        \centering
        \includegraphics[width=0.8\linewidth]{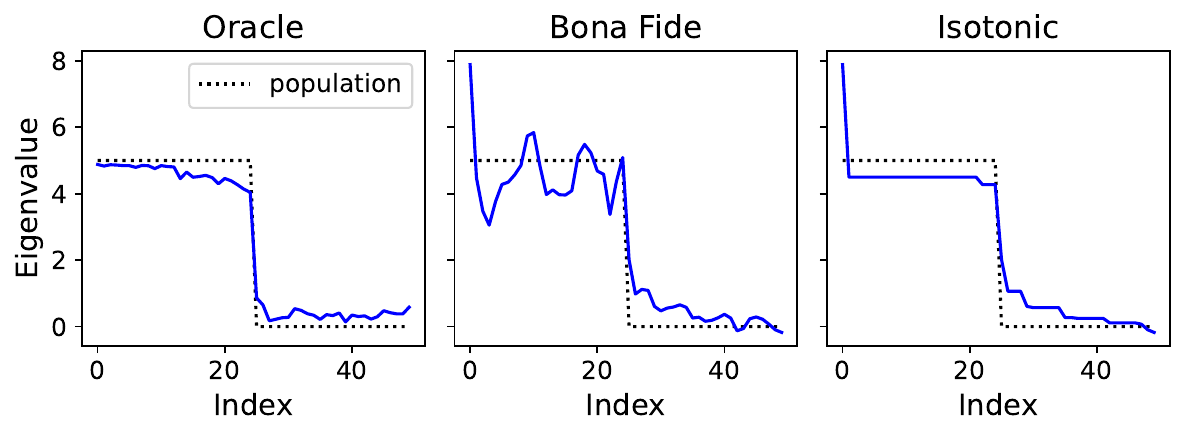}
        \caption{Step decay.}
    \end{subfigure}
    \caption{Bona fide estimators with and without isotonic post-processing under $\tau_{\max}=5$.}
    \label{fig:twub.bonafide}
\end{figure}

\begin{figure}[htb]
    \centering
    \begin{subfigure}[t]{\textwidth}
        \centering
        \includegraphics[width=0.8\linewidth]{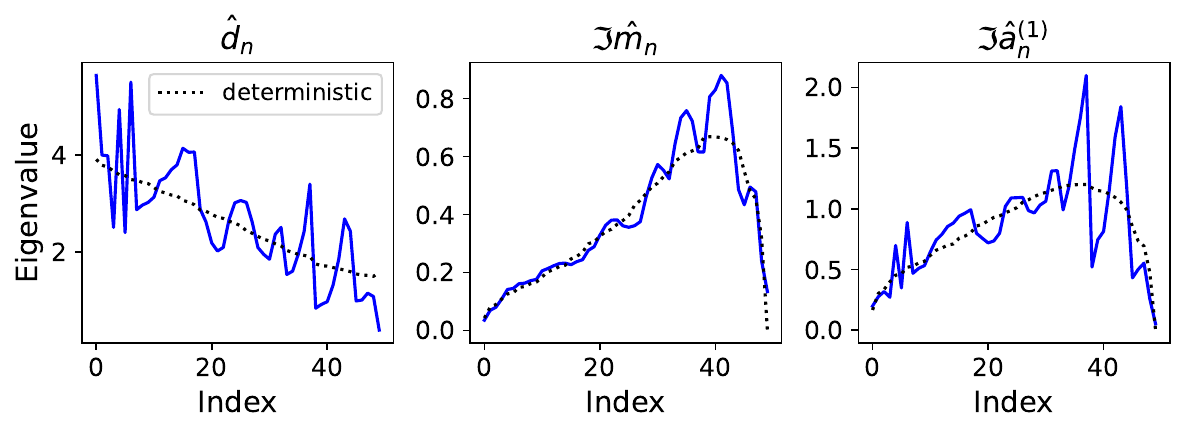}
        \caption{Linear decay.}
    \end{subfigure}
    \begin{subfigure}[t]{\textwidth}
        \centering
        \includegraphics[width=0.8\linewidth]{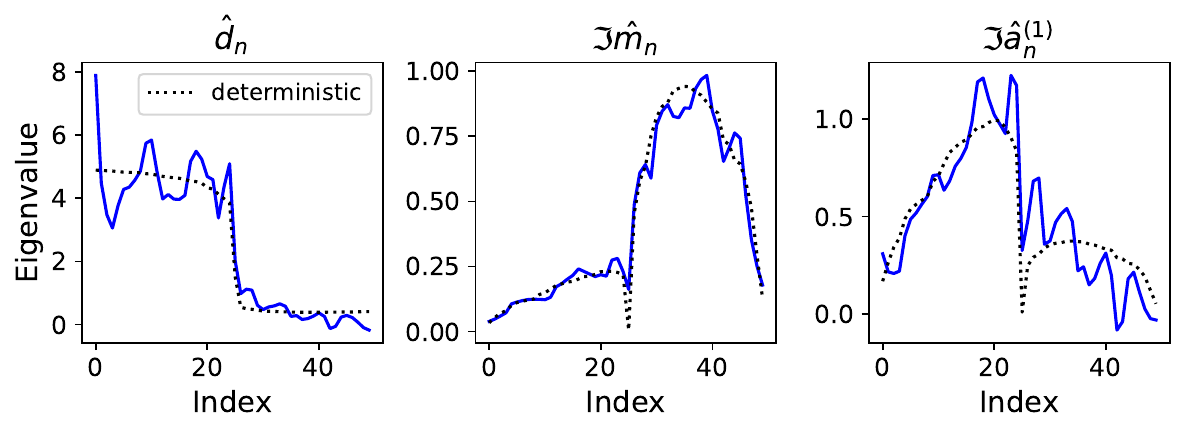}
        \caption{Step decay.}
    \end{subfigure}
    \caption{Bona fide estimators versus deterministic equivalent oracle estimators.}
    \label{fig:twub.deteqv}
\end{figure}

\paragraph{Bona fide nonparametric estimators.}
Figure~\ref{fig:twub.bonafide} compares the bona fide nonparametric estimates with their oracle counterparts, with and without isotonic post-processing. Estimation becomes less stable as the number of randomness levels to be deconvolved increases, and we observe that isotonic regression mitigates this instability. It is also worth noting that since the self-consistent system for $\{\hat a_r\}$ is solved numerically via root finding with random initializations, solutions can vary slightly for the same $\{\hat b_r\}$. Figure~\ref{fig:twub.deteqv} compares the deterministic-equivalent oracle estimator with the bona fide estimates, and likewise compare $\Im[\check a_n^{(1)}]$ and $\Im[\check m_n]$, the numerator and denominator of the deterministic-equivalent oracle, with $\Im[\hat a_n^{(1)}]$ and $\Im[\hat m_n]$. We approximate $\check a_n^{(1)}(z)$ and $\check m_n(z)$ by $\tilde a_n^{(1)}(z+i\eta)$ and $\tilde m_n(z+i\eta)$ with $\eta=0.001$, using the iterative algorithm in \cite[Theorem~1.5]{fan2017eigenvalue}.

\end{document}